\documentclass[twocolumn]{aastex63}
\usepackage{amssymb}
\usepackage{amsmath}
\usepackage{color}
\usepackage{slantsc}
\usepackage{lineno}
\usepackage{csvsimple}
\usepackage{xcolor}

\received{TBD}
\revised{TBD}
\accepted{TBD}
\submitjournal{PSJ}
\shorttitle{Atomic metals in the 1996 apparition of comet Hyakutake}
\graphicspath{{./}{figures/}}
\begin{document}
\title{Atomic iron and nickel in the coma of C/1996 B2 (Hyakutake): production rates, emission mechanisms, and possible parents}

\correspondingauthor{Steven Bromley}
\email{sjb0068@auburn.edu}
\author[0000-0003-2110-8152]{S. J. Bromley}
\affiliation{Department of Physics,  Auburn University, Leach Science Center, Auburn, AL 36832, USA}
\author{B. Neff}
\affiliation{Department of Physics and Astronomy, Clemson University, Clemson, SC, 29630, USA}
\author[0000-0002-3822-6756]{S. D. Loch}
\affiliation{Department of Physics,  Auburn University, Leach Science Center, Auburn, AL 36832, USA}
\author[0000-0002-1626-5615]{J. P. Marler}
\affiliation{Department of Physics and Astronomy, Clemson University, Clemson, SC, 29630, USA}
\author[0000-0002-4309-6060]{J. Országh}
\affiliation{Department of Experimental Physics, Faculty of Mathematics, Physics and Informatics, Comenius University in Bratislava, Mlynská dolina F-2, 842 48 Bratislava; Slovak Republic}
\author[0000-0003-3321-1472]{K. Venkataramani}
\affiliation{Department of Physics,  Auburn University, Leach Science Center, Auburn, AL 36832, USA}
\author[0000-0002-2668-7248]{D. Bodewits}
\affiliation{Department of Physics,  Auburn University, Leach Science Center, Auburn, AL 36832, USA}
%
\begin{abstract}
\noindent {Two papers recently reported the detection of gaseous nickel and iron in the comae of over 20 comets from observations collected over two decades, including interstellar comet 2I/Borisov}. To evaluate the state of the laboratory data in support of these identifications, we re-analyzed archived spectra of comet C/1996 B2 (Hyakutake), one of the nearest and brightest comets of the last century, using a combined experimental and computational approach. We developed a new, many-level fluorescence model that indicates that the fluorescence emission of \ion{Fe}{1} and  \ion{Ni}{1} vary greatly with heliocentric velocity. Combining this model with laboratory spectra of an Fe-Ni plasma, we  identified {22} lines of \ion{Fe}{1} and 14 lines of \ion{Ni}{1} in the spectrum of Hyakutake. Using Haser models, we estimate the nickel and iron production rates as $Q_\textrm{Ni} = 2.6 - 4.1\times10^{22}$~s$^{-1}$ and $Q_\textrm{Fe} = 0.4 - 2.8\times10^{23}$~s$^{-1}$. From derived column densities, the Ni/Fe abundance ratio {log$_{10}$[Ni/Fe] = $-0.15\pm0.07$} deviates significantly from solar abundance ratios, and it is consistent with the ratios observed in solar system comets. Possible production and emission mechanisms are analyzed in context of existing laboratory measurements. Based on the observed spatial distributions, excellent fluorescence model agreement, and Ni/Fe ratio, our findings support an origin consisting of {a short-lived} unknown parent followed by fluorescence emission. Our {models} suggest that the strong heliocentric velocity dependence of the fluorescence efficiencies can provide a meaningful test of the physical process responsible for the \ion{Fe}{1} and \ion{Ni}{1} emission.
\end{abstract}
\section{Introduction} \label{sec:intro}
Comets are the remnants of a protoplanetary disk, formed by accretion and preserved in the far reaches of our solar system. Upon transit towards the sun, solar irradiation of the surface sublimates the ices, liberating gas and dust into an expanding cloud known as the coma. Whereas the properties of the nucleus cannot be sampled directly by remote observations, atoms and molecules in the coma enable the study of both the atomic and molecular vestiges of the early solar system. These molecules primarily consist of H, C, N, S, and O in various formulations~\citep{Altwegg2019}. The bulk ices in the nucleus are H$_2$O, CO, CO$_2$, mixed with other trace species, some of which may reside primarily in the dust~\citep{Rubin2019}. Solar radiation photodissociates and photoionizes the molecules, leading to many radicals and atomic fragments \citep{Feldman2004}. Fragment species have many bright emission features in the visible and near-ultraviolet wavelengths, making them an easily accessible proxy to parent species~(cf.~\citealt{Ahearn1995}).

Thus far, the molecular inventory of comets numbers in the dozens, ranging from $\sim$30 parent species visible from ground-based observations~\citep{Bockelee2017} and doubling to $\sim$56 total \textit{in situ}~(cf.~67P/Churyumov-Gerasimenko, \citealt{Altwegg2019}). The detection of molecules in cometary comae sheds light on the abundances and trace organic chemistry during solar system formation.

Metal atoms, on the other hand, are mostly present in refractory materials. They have been discovered in e.g. the sample return mission~\textit{Stardust}, which collected dust from comet 81P/Wild~2 and revealed the presence of, among others, copper (Cu), iron (Fe), nickel (Ni), and zirconium (Zr) contained in olivine, sulfides, and other mineral structures~\citep{Berger2011,Brownlee2014}. Ca, Fe, K, Li, and Na which are normally absent in Jupiter's atmosphere were detected after the impact of D/1993 F2 (Shoemaker-Levy 9)~\citep{Roos1995}. Neutral and ionized Mg and Fe were detected in UV spectroscopy through meteorites in Mars' atmosphere after the close encounter with C/2013 A1 (Siding Spring) in Oct. 2014~\citep{Crismani2018}. Neutral sodium (\ion{Na}{1}) has been detected in several comets, including the tail of comet C/1995 O1 (Hale-Bopp)~\citep{Cremonese2002}, C/1996 B2 (Hyakutake) \citep{Wyckoff1999}, and most recently C/2020 F3 (NEOWISE). These detections summarize the limitations of detecting atomic metals in comets where these endeavors often require extreme events such as close approaches or direct collisions that liberate atoms from within the bulk composition of the nucleus.

In the absence of extreme conditions, detections are even rarer, being limited to sungrazing comets which traverse close to the sun~\citep{Jones2017}. At one to several AU, temperatures ($<300$~K) are insufficient for sublimating refractories or sulfides which require $T\geq680$~K~\citep{Pollack1994}. Sungrazers experience temperatures sufficient for sublimating metal-containing minerals such as olivine (Mg,Fe)$_2$SiO$_4$ that are otherwise inert at the 1 -- 3~AU distances common to most spectroscopic studies. The most famous sungrazer to-date was Comet C/1965 S1 (Ikeya-Seki), which passed the sun as close as 1.7~R$_{\odot}$~(0.008~AU) in 1965. The analyses by \cite{Preston1967} at 13~$R_{\odot}$ and \cite{Slaughter1969} at $\sim$30~$R_{\odot}$ revealed lines of \ion{Ca}{1}, \ion{Ca}{2}, \ion{Co}{1}, \ion{Cu}{1}, \ion{Fe}{1}, \ion{K}{1}, \ion{Mn}{1}, \ion{Na}{1}, \ion{Ni}{1}, and \ion{V}{1}, from which an effective excitation of 4480~K was derived, suggesting that metal atoms in the coma were excited by fluorescence. Thousands of sungrazers are known, but most are very small, too faint, and/or approach the Sun too closely for spectroscopy~\citep{Jones2017}.

Most recently, \cite{Manfroid2021} reported detections of (gaseous) \ion{Fe}{1} and \ion{Ni}{1} in the ultraviolet-visible range from a large sample of solar system comets observed with  ESO's Very Large Telescope. Rather fortuitously, the available compilations of atomic data for \ion{Fe}{1} and \ion{Ni}{1} in the NIST Atomic Spectra Database (hereafter denoted ASD,~\citealt{NIST_ASD}) are considerable, allowing for development of a detailed many-level atomic fluorescence model. Assuming the emission is driven by fluorescence, \cite{Manfroid2021} derived production rates and abundances of both iron and nickel from dozens of emission lines\footnote{Line list available as supplementary information at \url{https://www.researchsquare.com/article/rs-101492/v1}}. While the production rates of these species are low ($\sim$~grams/sec), the large fluorescence efficiencies enabled their detection.

The molecular parents of these metal atoms are currently unknown. As proposed by \cite{Manfroid2021}, one possibility is iron and nickel carbonyls, i.e. Fe(CO)$_5$ and Ni(CO)$_4$, whose sublimation temperatures are comparable to that of CO~\citep{Manfroid2021}, which would explain how metallic emission was observed even at large heliocentric distances up to 3.5~AU. Interestingly, these detections are not limited to solar system objects; the interstellar object 2I/Borisov was recently observed by \cite{Guzik2021} at 2.3~AU, revealing 9 atomic Ni emission features in the same wavelength window as \cite{Manfroid2021}. Taken together, it appears that the parents of both metals may be ubiquitous in cometary nuclei and suggest an as-yet-unexplored aspect of organic chemistry (cf. \citealt{Klotz1996}).

In this work, we present a new open-source model of the fluorescence emission of nickel and iron atoms based on atomic data in the ASD. We compare these models to laboratory data of \ion{Fe}{1} and \ion{Ni}{1} that were acquired as a background measurement for plasma experiments, and archived spectra of C/1996 B2 (Hyakutake) to confirm the line assignments of \cite{Manfroid2021} and \cite{Guzik2021}, add to the inventory of metal lines present in regular cometary spectra, and to evaluate their excitation and formation mechanisms. The fluorescence model will be made publicly available to aid future observers.

The remainder of this work is as follows. In Sec.~\ref{sec:hyak}, the laboratory spectra of iron and nickel is discussed and an overview of the archived spectra of C/1996 B2 (Hyakutake) is provided. In Sec.~\ref{sec:model}, the development of a fluorescence model utilizing the full extent of the available laboratory data is discussed. We detail the software implementation of our model, evaluate the propagation of the uncertainties of atomic data through to the {fluorescence spectrum}, and benchmark our model against spectra of C/1965~S1 (Ikeya-Seki). In Sec.~\ref{sec:results}, observed lines of \ion{Ni}{1} and \ion{Fe}{1} are discussed with special attention on the validity of the identifications. Lastly, in Sec.~\ref{sec:discussion} we compare our production rates and Ni/Fe abundances to previous works and elaborate on complexities of the parent identification.

\section{Data} \label{sec:hyak}
\subsection{Experimental data}\label{subsec:experiment}
The laboratory spectra utilized here were collected during the measurement campaign of Au~I and Au~II emission reported in \cite{Bromley2020}. These experiments used multiple probes containing stainless steel nickel-plated and gold-plated probes from which spectra of gold-iron-nickel plasmas and nickel-iron plasmas were collected. In short, the spectra were collected at resolving power {$\sim5\times10^{3}$} between 200 and 800 nm from erosion of the probes inside the Compact Toroidal Hybrid plasma apparatus at Auburn University. A description of the apparatus and optical scheme are available in \cite{Hartwell2017} and \cite{Johnson2019}. The spectra acquired in the lab lack emission from the `contaminants'  OH, CO, CN, and other volatiles that emit within the 300 - 500~nm window of the comet spectra. However, the lab spectra also show features of \ion{H}{1} and neutral and ionized C, N, and O. A detailed analysis of the Ni~I and Ni~II emission in the lab data will be reported in a future work (In Preparation). Though the excitation mechanisms of our plasma experiment are obviously different compared to a comet (electron impact versus photofluorescence), the lab spectra confirm the positions and detections of Ni~I emission lines. Electron impact excitation also excites different levels than photofluorescence, allowing us to look for previously undocumented transitions (cf. \citealt{Bodewits2019}).

\subsection{Observations}
The details of the observations, data reduction, and previous analyses of the C/1996 B2 (Hyakutake) spectra are discussed in \cite{Meier1998}, \cite{Kim2003}, \cite{AHearn2013}, and \cite{Ahearn2015}, and the data is available through NASA's Planetary Data System \citep{AhearnPDS2015}. In summary, the observations  of Comet C/1996 B2 (Hyakutake) were acquired on March 26, 1996  using the Echelle Spectrograph on the 4m Mayall Telescope at Kitt Peak National Observatory. The comet was observed with a slit size of 0.87$\times$7.4~arcsec where spectra were recorded in the range 300 -- 500~nm with resolving power $\sim$18,000{~\citep{Meier1998}}. At the time of observation, the comet was traveling at -36.7~km~s$^{-1}$ with respect to the sun at a heliocentric distance of 1.02~AU, and only 0.11~AU from Earth. The spectra were wavelength and intensity calibrated using Th-Ar lamps at each echellon order (42 total). The archived spectra were corrected for the wavelength dependent atmospheric extinction and \cite{AhearnPDS2015} used spectra of
solar analog 16~Cyg~B and `sky flats'  to remove the continuum caused by sunlight reflected by cometary dust.  Wavelength calibrations are accurate to $\pm0.02$~nm, and the (relative) intensity calibration within each order is estimated at the 5\% level~\citep{Meier1998}. However, overfilling of the slit during observations of solar analogs introduces an estimated $\pm40\%$ uncertainty on the absolute intensity calibration across the entire wavelength range.

\section{Fluorescence Model} \label{sec:model}
While our laboratory spectra are useful for reporting line positions, the relative line heights cannot be directly compared to {a} comet {spectrum} as the excitation mechanisms are fundamentally different. In {a} lab plasma, the {spectrum varies} as a function of the electron temperature and density. For a fluorescing media, the {spectrum} is determined by the interplay of absorption, stimulated emission, and spontaneous emission, whose equilibrium balance will change with both heliocentric velocity and distance.

Using a fluorescence model, comparisons of synthetic versus observed spectra reveal clues on the physical process of atomic metal emission in the coma, as well as the chemical origins of these atoms. {For example, fluorescence is unlikely to populate highly excited states, which are more likely to be excited from prompt emission.} While one could assume a two-level model to study individual lines driven by absorption from the ground state, atoms with complex electronic structures require a many-level model to properly account for cascades and excited states populated by multiple absorption pathways.

We developed a fluorescence model assuming a collisionless and optically thin environment for the emitted photons, driven by three processes: spontaneous emission~(1), stimulated emission (2), and absorption (3). While the rate for (1) is a fundamental property, the rates for (2) and (3) derive from the rate of (1) and the local radiation field at the comet. In the following we outline our fluorescence model; the presentation of the matrices and differential equations is limited to 2 levels for readability.

Consider two levels of a many-level system with a transition from upper level $j$ to lower level $i$. If the Einstein $A$ coefficient for (1), $A_{j\rightarrow{i}}$, is known, the coefficients for (2) and (3) may be expressed as
\begin{equation}\label{eq:coeff_stim}
B_{j\rightarrow{i}} = \frac{\lambda_{ji}^5}{8\pi{hc^2}}A_{j\rightarrow{i}}
\end{equation}
\begin{equation}\label{eq:coeff_ab}
B_{i\rightarrow{j}} = \frac{g_j}{g_i}B_{j\rightarrow{i}}
\end{equation}
where $\lambda_{ji}$ is the (vacuum) wavelength of the transition, $g_j = 2J_j + 1$, and $g_i = 2J_i + 1$. For each $B$, the units [m$^3$~W$^{-1}$~s$^{-1}$] ensure compatibility with the choice of solar spectrum described later. For a given pair of levels $j$ and $i$, the change in population of the upper state $j$ is written
\begin{equation}\label{eq:dnj}
\begin{split}
\frac{dn_j}{dt} = -A_{j\rightarrow{i}}n_j - B_{j\rightarrow{i}}n_j{\int^{\lambda+\Delta\lambda}_{\lambda-\Delta\lambda}\phi(\lambda_{ij})\rho(\lambda_{ij})d\lambda} \\
+ B_{i\rightarrow{j}}n_i{\int^{\lambda+\Delta\lambda}_{\lambda-\Delta\lambda}\phi(\lambda_{ij})\rho(\lambda_{ij})d\lambda}
\end{split}
\end{equation}
with $B_{ji}$ and $B_{ij}$ from Eqn's.~\ref{eq:coeff_stim} and \ref{eq:coeff_ab}, $\rho(\lambda)$ is the flux per wavelength interval (W~m$^{-3}$) incident on the population $n_j$ at the energy of the transition from level $j$ to level $i${, and $\phi(\lambda_{ij})$ is the line shape (discussed in Sec.~\ref{sec:python})}. The population of the lower state, $n_i$, may be written similarly as
\begin{equation}
\begin{split}
\frac{dn_i}{dt} = A_{j\rightarrow{i}}n_j + B_{j\rightarrow{i}}n_j{\int^{\lambda+\Delta\lambda}_{\lambda-\Delta\lambda}\phi(\lambda_{ij})\rho(\lambda_{ij})d\lambda}\\ - B_{i\rightarrow{j}}n_i{\int^{\lambda+\Delta\lambda}_{\lambda-\Delta\lambda}\phi(\lambda_{ij})\rho(\lambda_{ij})d\lambda}
\end{split}
\end{equation}

{Abbreviating $\sigma_{ij} \equiv {\int^{\lambda+\Delta\lambda}_{\lambda-\Delta\lambda}\phi(\lambda_{ij})\rho(\lambda_{ij})d\lambda}$,} the above system of equations for levels $i$ and $j$ may be written in matrix form as
\begin{equation}\label{eq:2lev}
\begin{bmatrix}
-{\sigma_{ij}}B_{i\rightarrow{j}} & A_{j\rightarrow{i}}+ {\sigma_{ij}}B_{j\rightarrow{i}} \\
{\sigma_{ij}}B_{i\rightarrow{j}}\ & -A_{j\rightarrow{i}} - {\sigma_{ij}}B_{j\rightarrow{i}}
\end{bmatrix}
\begin{bmatrix}
n_i \\
n_j
\end{bmatrix}
=
\begin{bmatrix}
dn_i/dt \\
dn_j/dt
\end{bmatrix}
\end{equation}

In equilibrium, the populations are constant in time and thus the right hand side of Eq.~\ref{eq:2lev} is equal to 0. However, the matrix $A$ is underdetermined, i.e. $N$ equations and $N-1$ unknowns. We add an additional constraint by enforcing a normalization condition, $\sum_{i}n_i = 1$, by replacing all elements in row 0 of both the left and right matrices by 1. Our matrix equation thus takes the form

\begin{equation}\label{eq:matrix_form}
\textbf{A}\vec{x} = \textbf{B}
\end{equation}
where the rates for processes (1) -- (3) are stored in matrix $\textbf{A}$ with populations contained within column vector $\vec{x}$. {For each pair of levels in the model, the} transition rate contributes to two matrix elements: a positive contribution to the off-diagonal element $(i,j)$, and a negative contribution to the diagonal element $(j,j)$. Similarly, stimulated emission contributes positively to element $(i,j)$ and negatively to element $(j,j)$. Absorption provides a negative contribution to the diagonal term of level $i$ at $(i,i)$, and an off-diagonal, positive contribution to the population of $n_j$ in element $(j,i)$.

After populating matrices $A$ and $B$, the equilibrium population fractions $\vec{x}$ follow as $\vec{x} = A^{-1}\times{B}$. The transition intensity, i.e. the fluorescence efficiency {(`g-factor')} of the transition {in units of J~s$^{-1}$~particle$^{-1}$} \textit{at the emitting source} from level $j$ to level $i$ then follows from the level population $n_j$ as
\begin{equation}\label{eq:intens}
I_{j\rightarrow{i}} = \frac{hc}{\lambda_{ji}}n_j{A_{j\rightarrow{i}}}
\end{equation}
where given the directionality of stimulated emission along the sun-comet vector the contribution of stimulated emission to the observed line intensities is assumed to be negligible.


\subsection{Computational Implementation}\label{sec:python}
We implemented the fluorescence model in Sec.~\ref{sec:model} in Python3. The code requires only standard Python packages, and performs all mathematical operations using NumPy~\citep{numpy} and SciPy\cite{scipy}. SI units are utilized throughout, with conversions indicated where necessary.  The code is publicly available at the authors GitHub page\footnote{\url{https://github.com/StevenBromley/fluorescence_model}}.

First, Einstein $A$ coefficients and level information (energies, $J$ values) were retrieved from the {National Institute of Standards and Technology (NIST) Atomic Spectra Database (ASD,~\citealt{NIST_ASD}). From the level information, we generate wavelengths in vacuum (`Ritz' wavelengths), and convert to air wavelengths after all computations are carried out.} For each transition, stimulated emission and absorption coefficients follow from Eqn's~\ref{eq:coeff_stim} and \ref{eq:coeff_ab}. To generate absorption and stimulated emission \textit{rates}, {we compiled a high-resolution solar spectrum from several space- and ground-based observations around solar minimum.} Our (vacuum) flux atlas was therefore compiled as follows:

{
\begin{itemize}
    \item Between 0 -- 168 nm, we utilize the coronal/chromosphere far ultraviolet solar model `13x1' of \cite{Fontenla2014}.
    \item Between 168 -- 200.06 nm, we assumed the solar {spectrum} collected during solar minimum from the SOLSTICE instrument with 0.1~nm resolution.\footnote{Available for download at \url{https://lasp.colorado.edu/home/sorce/data/ssi-data/}}
    \item For the small region 200.06 - 202 nm, data from the \cite{AH1991} dataset was chosen to minimize discontinuities at the edges between adjacent datasets.
    \item Between 202 -- 2730~nm, we assumed the calibrated solar spectrum from \cite{Coddington2021}, which itself was compiled from several high-resolution solar datasets, including the observations of \cite{Kurucz1984}.
\end{itemize}
}

{For each application the flux per wavelength interval $\rho(\lambda)$ is re-scaled to the comet's heliocentric distance by a factor of $1/r_{\textrm{h}}^2$. By default, our combined flux atlas is scaled to 1 AU and will be made available in several forms (e.g. with/without the FUV components) with our model code. The flux atlas is continuous at nearly all boundaries and integrates to 1327 W~m$^{-2}$ at 1~AU. The `missing' 3\% flux, compared to a nominal integrated solar flux of $\sim$1370 W~m$^{-2}$, arises from the infrared longward of 2730~nm.  For wavelengths outside the bounds of the {assumed} solar {spectrum}, the code is capable of assuming a blackbody radiation field, defaulted to 5777~K.}

If desired, the user is free to import and utilize any radiation field(s), such as those around stars other than the sun for applications such as exocomet studies~\citep{Strom2020} or handling fluorescence emission of \ion{Fe}{2} in active galactic nuclei~\citep{Nierenberg2019,Kova2010}. {We note that our choice of solar spectrum is not applicable for solar conditions far from the conditions of solar minimum, particularly in the ultraviolet portion of the solar spectrum which varies strongly with solar activity.}

We tested the sensitivity of our model to the choice of solar spectrum by comparing {fluorescence spectra generated using our compiled solar spectrum against those generated using the lower-resolution solar atlas spanning 200 -- 1001~nm reported in \cite{Chance2010}. On average, g-factors calculated with this solar spectrum differ by $\sim$10\% for both \ion{Ni}{1} and \ion{Fe}{1}. In the range 300 -- 400~nm, g-factors of \ion{Ni}{1} are relatively insensitive at $v_\textrm{h} = -36.7$~km~s$^{-1}$, though some g-factors of \ion{Fe}{1} show relative differences as high as 28\%.} The lower resolution of this solar spectrum (FWHM 0.5~nm versus $\sim$0.002~nm above) makes such a choice likely unsuitable for this application, and {we use our compiled high-resolution solar spectrum for the remainder of this work.}

When calculating the absorption and stimulated emission rates, {$W = B\int\phi(\lambda)\rho(\lambda)d\lambda$}, the Doppler shift of the solar spectrum resulting from the comet's heliocentric velocity (-36.7 km/s during the archived observations of Hyakutake) {is included by shifting the line center and line profile. For the line profile, two choices are provided: a delta function, and a Doppler (thermal) line profile. The Doppler profile width is taken as
\begin{equation}
    \sigma_\textrm{doppler} = \lambda_{ij}\sqrt{\frac{2k_bT}{mc^2}}
\end{equation}
where $c$ is the speed of light, $k_b$ is the Boltzmann constant, $T$ = $279~r_h^{-1/2}$ with $r_h$ in AU is the assumed blackbody temperature of the comet, and $m$ is the mass of the emitting particle. For each line, the profile $\phi(\lambda_{ij})$ is normalized such that $\int\phi(\lambda)d\lambda = 1$ and is presented numerically on a (default) 500 point grid spanning $\lambda_{ij}\pm0.005$~nm. For models used to analyze the spectrum of C/1996 B2 (Hyakutake), Doppler line profiles with $T = 276$~K are assumed.
}

After the initial matrix population, the inverse of the rate matrix ($A^{-1}$) is calculated using the NumPy \textbf{linalg.inv} package, from which the populations follow as $\vec{x} = A^{-1}\times{B}$. For large singular or near-singular matrices caused by missing atomic data or small heliocentric distances (i.e. large absorption rates), the code attempts a solution using a Psuedo-Inverse~(\textbf{linalg.pinv}) and/or Singular Value Decomposition (\textbf{linalg.svd}). From the equilibrium populations, {g-factors} are calculated using Eq.~\ref{eq:intens}, and vacuum wavelengths are Doppler-shifted by the geocentric comet velocity before conversion to standard air wavelengths using the conversion of~\cite{Morton2000}. We note that for the  observation of C/1996 B2 (Hyakutake) utilized here the geocentric velocity was effectively 0~km~s$^{-1}$, and the observation was recorded {at} $\Delta = 0.11$~AU.

Lastly, our model code directly imports the line and level lists which are downloaded from the ASD, and can thus be used to analyze any neutral or ionized atomic species for which the requisite data is available {and all included levels are connected (directly or indirectly) to the ground state}.

\begin{table}[t]
\centering
\begin{tabular}{lccc}
\centering
{Flag} & {Uncertainty(\%)} & \# Ni & \# Fe \\
\hline
AAA & $\leq$ 0.3 & 0  & 0\\
AA & $\leq$ 1 & 0 & 0 \\
A+ & $\leq$ 2 & 0 & 0 \\
A & $\leq$ 3 & 0 & 149\\
B+ & $\leq$ 7 & 0 & 282\\
B & $\leq$ 10 & 0 & 622\\
C+ & $\leq$ 18 & 56 & 509\\
C & $\leq$ 25  & 126 & 205\\
D+ & $\leq$ 40  & 45 & 188 \\
D & $\leq$ 50  & 211 & 191\\
E & $>$ 50 & 84 & 393\\
\hline
\end{tabular}
\caption{Einstein $A$ value accuracy scale in the NIST Atomic Spectra Database~\citep{NIST_ASD}. The number of lines with each rating for \ion{Ni}{1} and \ion{Fe}{1} are shown.}\label{tab:accur}.
\end{table}
%

\subsection{Error Propagation}\label{sec:error}
The Einstein $A$ values were sourced from the ASD~\citep{NIST_ASD} in which each $A$ value is assigned an accuracy rating. The ASD $A$ value accuracy scale is shown in Table~\ref{tab:accur}, and spans from $\leq~0.3$\% to $>$~50\%. Though some $A$ values used in our model may have large uncertainties in excess of 50\%, line intensities from the model cannot simply be attributed an uncertainty equal to that of the $A$ value. For each transition, the 6 resulting contributions to the rate matrix each modify the detailed balance which determines the equilibrium populations. Therefore, we adopted the following Monte-Carlo procedure for approximating the \textit{uncorrelated} uncertainties of the model intensities from the $A$ value uncertainties.

\begin{figure*}[!t]
\centering
\includegraphics[width=0.95\textwidth]{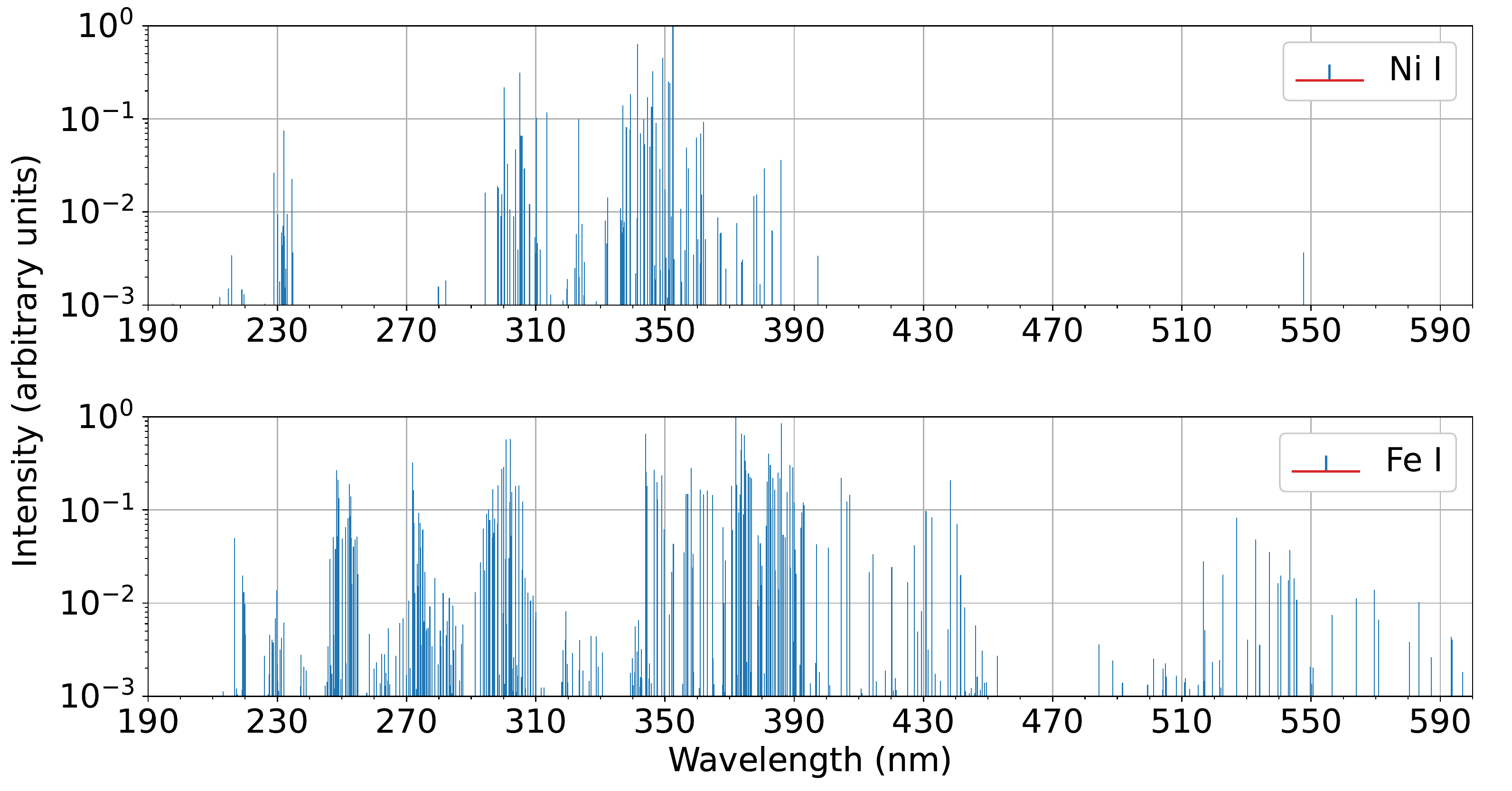}
\caption{Synthetic fluorescence spectra {stick plots} on a logarithmic intensity scale with the strongest lines normalized to 1 for \ion{Ni}{1} (top) and \ion{Fe}{1} (bottom) spanning 190 - 600~nm. Line intensities were calculated for the observing conditions of C/1996 B2 Hyakutake, $v_{\textrm{h}} = -36.7$~km/s and $r_{\textrm{h}} = 1.02$~AU.}\label{fig:synthetic_spec}
\end{figure*}

\begin{figure}
\centering
\includegraphics[width=0.5\textwidth]{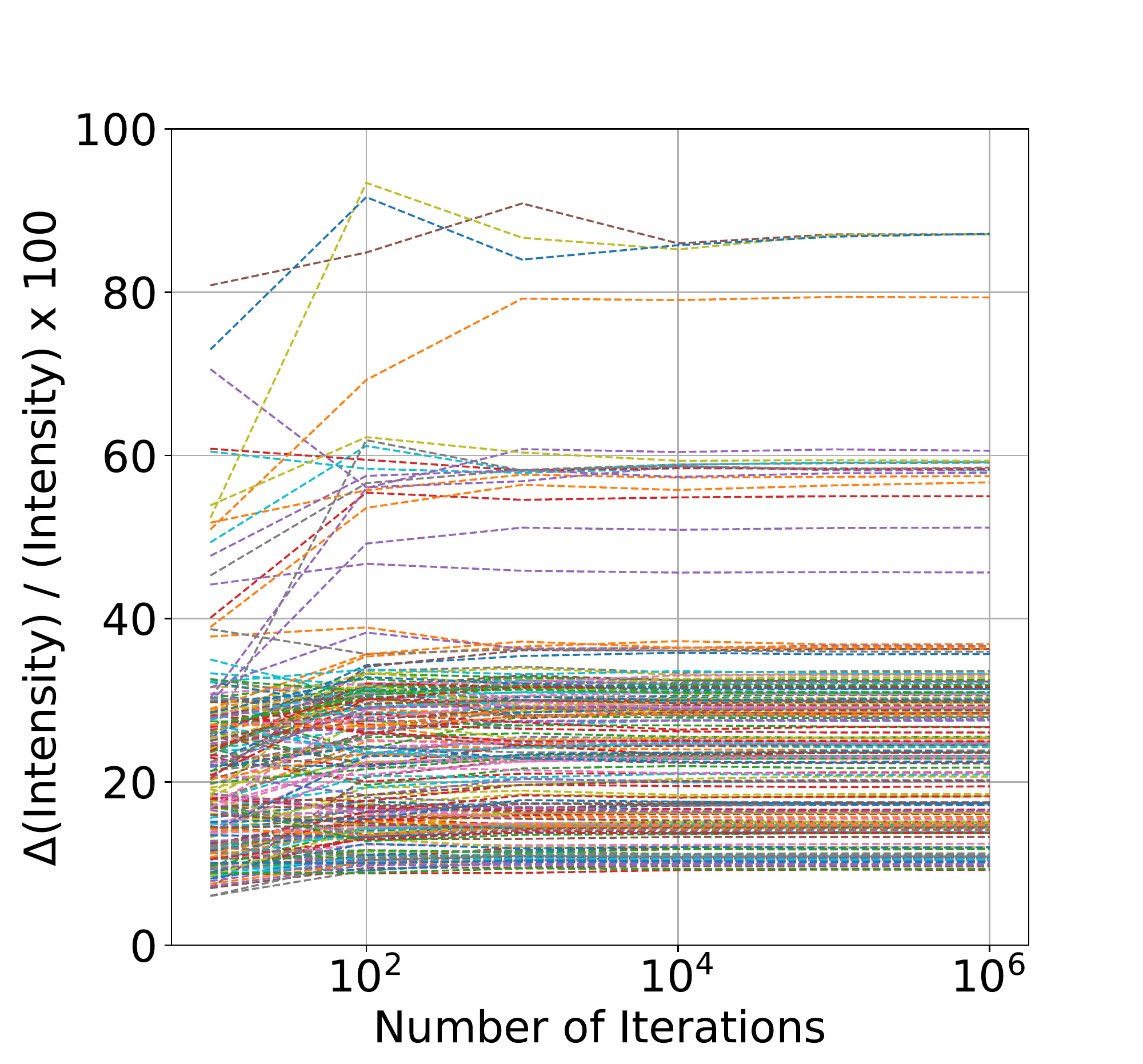}
\caption{Ratio of line intensity standard deviation to line intensity (in percentage) for 213 \ion{Ni}{1} lines spanning 300 - 500~nm as a function of iteration number (see text). Spectra were generated at the orbital conditions of the closest approach of C/1996 Hyakutake and are plotted at 10, 10$^2$, 10$^3$, 10$^4$, 10$^5$, and 10$^6$ iterations. Each colored line shows the convergence of an individual transition.}\label{fig:ni0_convergence}
\end{figure}

Error bars on model intensities are derived by looking at changes in line intensities following the simultaneous adjustment of all $A$ values. For each iteration, $A$ values are randomly and uniformly varied within a range of $\pm$ the accuracy rating e.g. $\pm$~10\% for `B' ratings. For the lowest accuracy rating (E, $>$50\%), we allow variations up to $\pm$100\%. We calculate the error bar for each emission line, $\Delta{I_{j\rightarrow{i}}}$, as the standard deviation of the line intensity across all iterations. We note that this is likely an overestimate of the uncorrelated uncertainty, as a Gaussian distribution (compared to the present uniformly distributed uncertainties) would favor lower variations for each $A$ value adjustment.

\subsection{\ion{Ni}{1} Model}\label{sec:ni_model}
For Ni~I, the 522 transitions in the ASD were incorporated into our fluorescence model, involving 133 of the 288 known levels. For the 155 excluded levels, no transitions with transition rates were available. A synthetic {spectrum} of Ni~I at 1.02~AU and a heliocentric velocity $v_{\textrm{h}} = -36.7$~km/s is shown in Fig.~\ref{fig:synthetic_spec}~$(top)$.The strongest emissions are between 300 - 390~nm. The model also suggests possibly-observable emission (from space-based missions) around $\sim$230~nm, but this emission is outside the bandpass of the archived {spectrum} of Hyakutake. The expected strongest emission overlaps with the wavelength ranges wherein nickel emission was identified by \cite{Manfroid2021} and \cite{Guzik2021}.

We derived error bars for the \ion{Ni}{1} {g-factors} by iterating our Monte-Carlo procedure with $10^n$~($n = 0 - 6$) iterations at the observing conditions of C/1996 B2 (Hyakutake). Fig.~\ref{fig:ni0_convergence} shows the convergence of our method with increasing number of iterations; only lines in the range 300 -- 500~nm are shown for visibility. The majority of the standard deviations converge by $10^4$ iterations, which requires $\sim$20~minutes of compute time on a modern personal computer.

\begin{figure}
\centering
\includegraphics[width=0.5\textwidth]{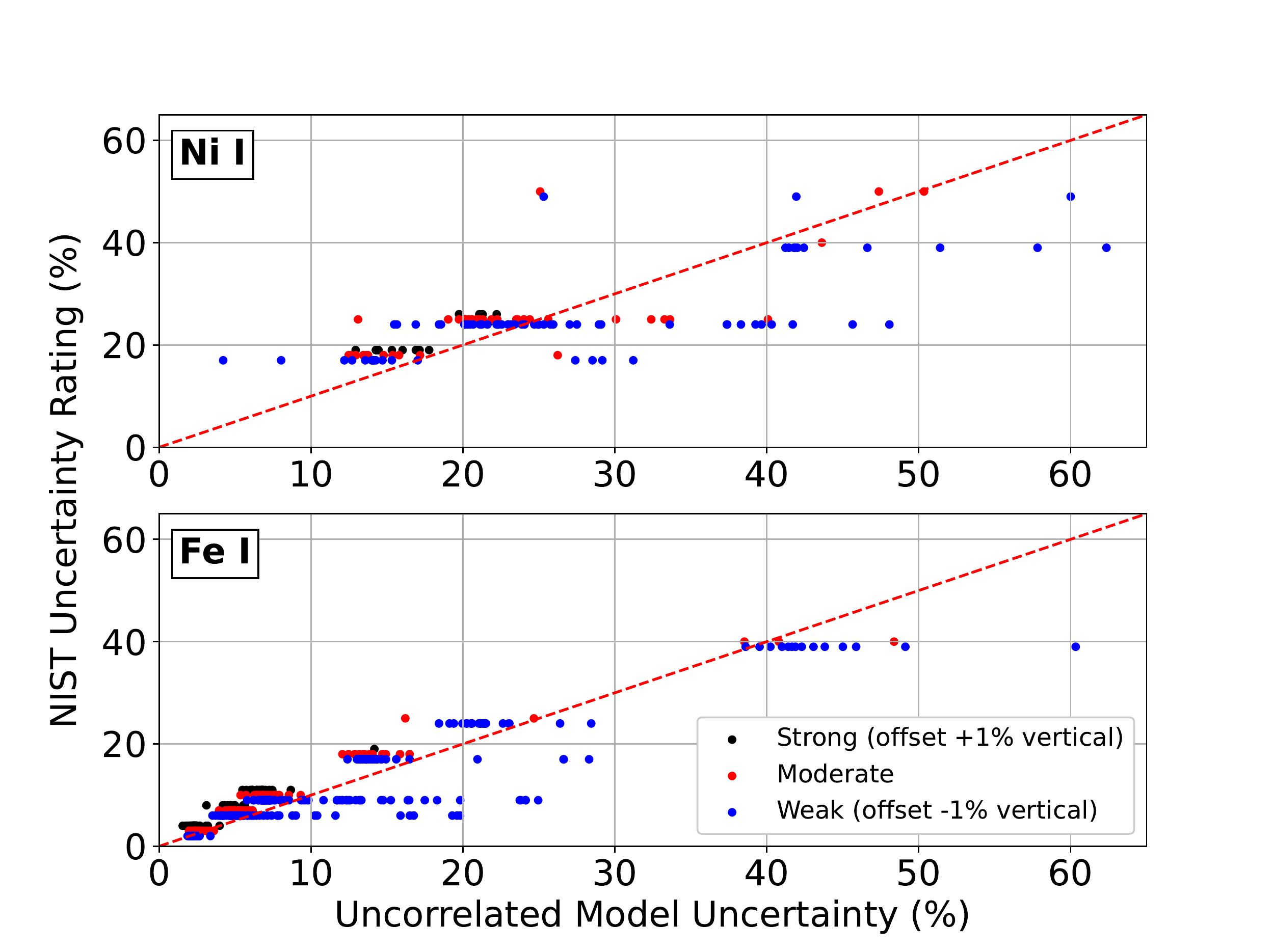}
\caption{\textit{(Top)} Model uncertainties for the fluorescence spectrum of \ion{Ni}{1}, calculated as the standard deviation of line intensities following $10^6$ iterations within the limits of $\pm$~ASD Accuracy (see Table~\ref{tab:accur}) versus the ASD accuracy rating for {134} selected transitions with model uncertainties $< 100\%$. {Axis limits were restricted to enhance visibility of the majority ($>99\%$) of lines within this criteria.} {Strong and weak lines are offset $\pm$1\% vertically with respect to moderate lines for visibility; for definitions of line classifications, see text.} \textit{(Bottom)} \ion{Fe}{1} model uncertainties versus ASD accuracy ratings for {425} select transitions with uncertainties lower than 100\%: {67} strong, {108} moderate, and {250} weak.}\label{fig:ni0_uncerts}
\end{figure}

Fig.~\ref{fig:ni0_uncerts} (top) shows the ratio of model standard deviation to model line intensity (in percentage) versus the NIST stated uncertainty rating (percentage) following $10^6$ iterations for selected lines of \ion{Ni}{1}. The dashed red line $y=x$ is plotted as a visual aid only. `Strong' lines (black) are those with a model uncertainty with an intensity within a factor of 10 of the strongest line. `Moderate' lines (red) are those with $10 < I_\textrm{strongest}/I_\textrm{line} < 100$, and `weak' lines (blue) are those with $100 < I_\textrm{strongest}/I_\textrm{line} < 1000$. A total of {134} lines meeting these criteria are shown.

The remaining (not shown) lines are very weak and exhibit large model uncertainties. We note that the 84 `E'-rated lines all present significant sensitivity to our Monte-Carlo procedure and exhibit extremely large uncertainties in excess of $10^{8}~\%$, owing to floating point errors and/or extremely small level populations. Thirty-nine of these 84 lines are E2 transitions between even-parity configurations where the upper levels are weakly populated by fluorescence; the dominant source of their population is absorption from the (already low) populations of excited odd parity states. Twenty-one separate `E'-rated transitions are from excited even configurations to the excited odd configurations, and suffer from a similar limitation. The remaining 24 `E'-rated transitions are outside our region of interest (300 -- 500~nm).

For weak lines, the small scales of both level populations and fluorescence efficiencies presents computational difficulties. For possible future investigations of particular lines using our model, it may be appropriate to investigate the model's sensitivity to changes in particular lines within some constraint e.g. the effect of branching ratio uncertainty for a particular upper level. However, such an investigation is outside the scope of this work.

For strong and moderately strong lines, the model uncertainty is comparable to or less than the stated ASD uncertainty rating. The reason for this is likely due to the following: recall that for a two-level system the populations follow as
\begin{equation}\label{eq:2lev_solution}
n_2 = n_1\frac{W_{1\rightarrow2}}{W_{2\rightarrow1} + A_{2\rightarrow1}}
\end{equation}
where $W_{1\rightarrow2}$ is the absorption rate, $W_{2\rightarrow1}$ is the stimulated emission rate, and $A_{2\rightarrow1}$ is the spontaneous emission rate. By definition, the Einstein B coefficients are proportional to the $A$ value, and the net effect of an increase in $A$, e.g. $A \rightarrow 2\times A$ propagates through the $B$ coefficients and leaves the ratio $n_2/n_1$ unchanged. However, the change in $A$ value propagates through the {g-factor} from Eq.~\ref{eq:intens} as $g_{2\rightarrow{1}} \propto A_{2\rightarrow{1}}$, and thus the {g-factor uncertainty is equal to the uncertainty of the transition rate.}

For a many-level system, we find that intensities for levels that are well-connected and well-populated are less affected by changes in $A$ values compared to levels connected by fewer transitions as the changes in multiple $A$ values partially cancel. For levels with one or few connections to the bulk of the levels, the uncertainty of the $A$ value more directly propagates through to the uncertainty of the line intensity. Given the scale of the \ion{Ni}{1} rate matrix, $133^2$ possible elements populated by $522\times6 = 3132$ rates, the $A$ value iterations for well-populated levels (which thus produce the strongest lines) are effectively a perturbation. In this case, the quantity of atomic data, assumed to accurately reflect the totality of the level structure, including all important transitions, acts as a buffer against large fluctuations in model outputs for the strongest lines.

\subsection{\ion{Fe}{1} Model}\label{sec:fe_model}
For Fe~I, we included the 2542 transitions with transition rates in the ASD, involving a total of 434 levels. Though more than 50\% of the transitions are outside our spectral window, they may affect the level populations and were included. Our flux atlas was sufficient for producing absorption and stimulated emission rates for all but {four M1 transitions between levels of the configuration $3d^64s^2$, which were approximated with a blackbody}.

Fig.~\ref{fig:synthetic_spec}~$(bottom)$ shows {a} synthetic \ion{Fe}{1} fluorescence {spectrum} at the observing conditions of comet Hyakutake \citep{Meier1998}. The strongest lines are predicted in the UV between $\sim$340 - 390~nm. For the \ion{Fe}{1} Monte-Carlo iterations, the error bars of most lines ($>99\%$) converge within $10^4$ iterations. In Fig.~\ref{fig:ni0_uncerts} (bottom), we show a comparison of our Monte-Carlo uncertainty calculation versus the NIST accuracy ratings of the transition rates for {425} selected lines following the procedure derived for \ion{Ni}{1}. Compared to \ion{Ni}{1}, most of the strongest lines of \ion{Fe}{1} exhibit a Monte-Carlo uncertainty less than the ASD accuracy rating. For weak lines, the Mont-Carlo uncertainty {is generally comparable to or larger than the ASD uncertainty rating.}

\subsection{Total Model Uncertainties}\label{subsec:total_uncert}

For the archived data of C/1996 B2 (Hyakutake), the heliocentric velocity (-36.7~km/s) leads to favorable fluorescence efficiencies. {We calculated the fluorescence efficiencies of \ion{Ni}{1} and \ion{Fe}{1} across a grid of heliocentric velocities spanning $v_{h}~\pm$1~km/s. {We consider the range of $\pm$1~km~s$^{-1}$ an upper bound, as this covers a range of typical outflow velocities in comae and is larger than the thermal speed of Ni and Fe atoms at 270~K (0.2 km~s$^{-1}$).}Within this range the average change in fluorescence efficiencies for all lines was {0.9\%} for \ion{Ni}{1} and {1.5\%} for \ion{Fe}{1}. Our total model uncertainties were thus obtained by adding in quadrature the uncertainty of the converged Monte-Carlo iterations and the above heliocentric velocity sensitivity.}
\begin{figure*}
\centering
\includegraphics[width=1\textwidth]{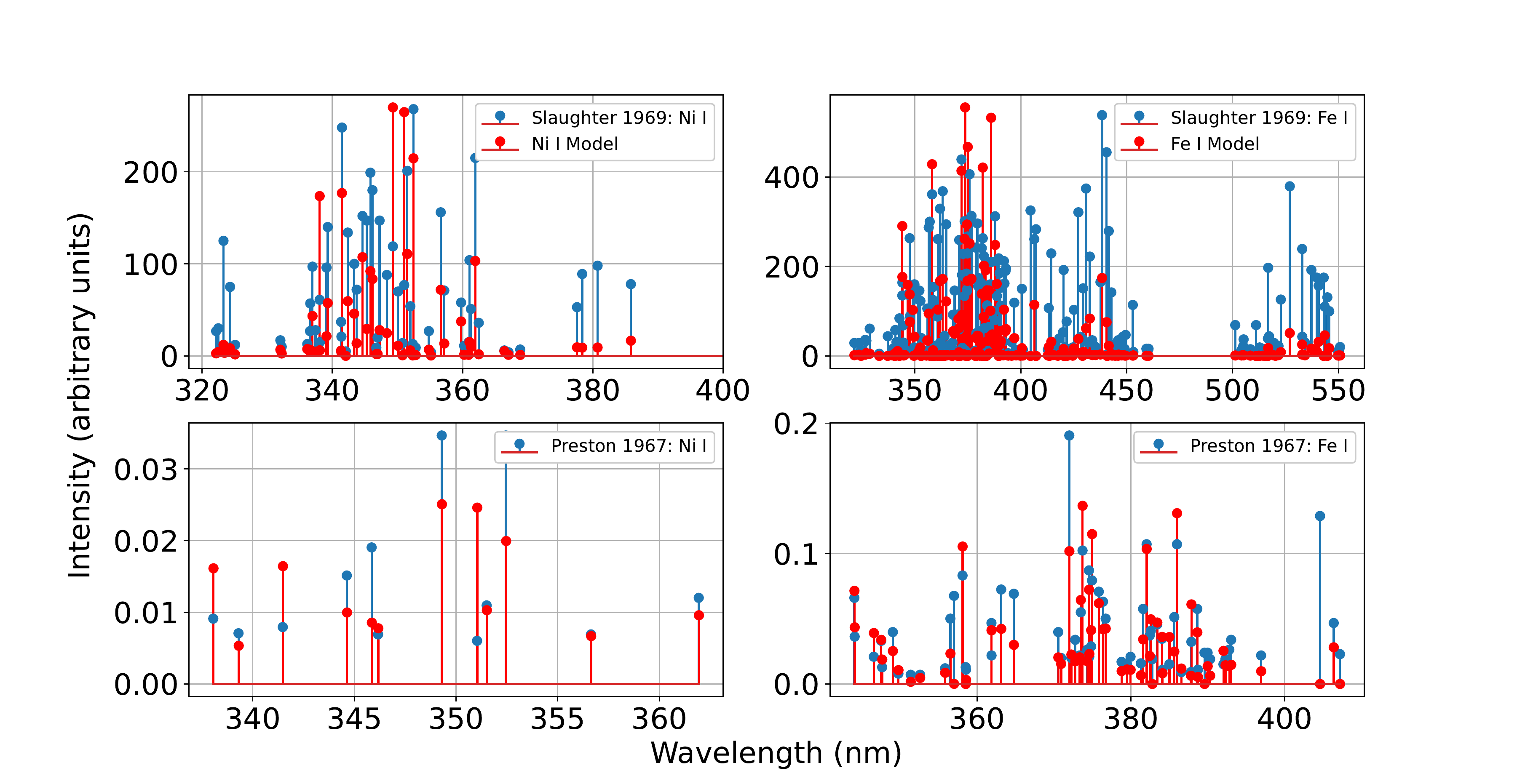}
\caption{Comparison of \ion{Ni}{1} (left) and \ion{Fe}{1} (right) line intensities of our fluorescence model to the observed intensities of comet C/1965 S1 (Ikeya-Seki) as reported in \cite{Slaughter1969} (top) and \cite{Preston1967} (bottom). The model spectra are scaled via a linear fit of all present lines. {Doppler line profiles were assumed for equilibrium temperatures derived from the heliocentric distances: 1067~K~\citep{Slaughter1969} and 818~K~\citep{Preston1967}.}}\label{fig:is_comp}
\end{figure*}


\subsection{Benchmarking Against Sungrazer Ikeya-Seki}\label{sec:IS}
As an additional check on the accuracy of the fluorescence model, we applied our model to {two} spectra of \ion{Fe}{1} and \ion{Ni}{1} collected during the perihelion passage of comet C/1965 S1 (Ikeya-Seki) in 1965. Two line lists are available from \cite{Slaughter1969} at 30~$R_{\odot}$ ($v_\textrm{h}\sim154$~km~s$^{-1}$) and \cite{Preston1967} at $\sim$13~$R_{\odot}$ ($v_\textrm{h}\sim116$~km~s$^{-1}$)\footnote{\url{https://ssd.jpl.nasa.gov/horizons.cgi}}.

Fig.~\ref{fig:is_comp} shows the comparison of our model spectra and intensities of \ion{Fe}{1} and \ion{Ni}{1} lines reported by \cite{Slaughter1969} and \cite{Preston1967}. The {modeled spectra were scaled by a linear fit of modeled versus observed line intensity for all metal lines.} The Ikeya-Seki spectra were recorded on photographic plates, and thus a direct comparison between the model versus observed intensities is difficult. \cite{Slaughter1969} notes that agreement between expected and observed line intensities of a thorium-argon lamp agreed within a factor of 2. \cite{Preston1967} reported intensities in terms of the local sky continuum, and thus we expect their intensity scale may differ as a function of wavelength. We assume a similar factor of 2 scatter due to intensity calibration.

The best agreement between our model and the reported lines in the spectra of Ikeya-Seki is found for the {spectrum} collected at 13~$R_{\odot}$ by \cite{Preston1967}. The agreement with all observations, particularly the {spectrum at} 30~$R_{\odot}$~\citep{Slaughter1969}, is improved significantly if a linear offset is included in the fit (assumed to account for reflected solar continuum). The large thermal temperature ($T > 4000~K$) in Ikeya-Seki may have introduced {additional excitation mechanisms}, but this effect would be more pronounced for the spectra of \cite{Preston1967}. The source of the disagreement between our model and the line list of \cite{Slaughter1969} is not known, {but it is likely that the strongest lines saturated the photographic plates, leading to a relative suppression of the weaker features in the scaling of our model to the observed line intensities}. However, lines that were strong in the Ikeya-Seki spectra have appreciable g-factors in our models, and we thus consider this agreement as sufficient validation of our model.

%
\begin{figure*}[t]
\centering
\includegraphics[width=1\textwidth]{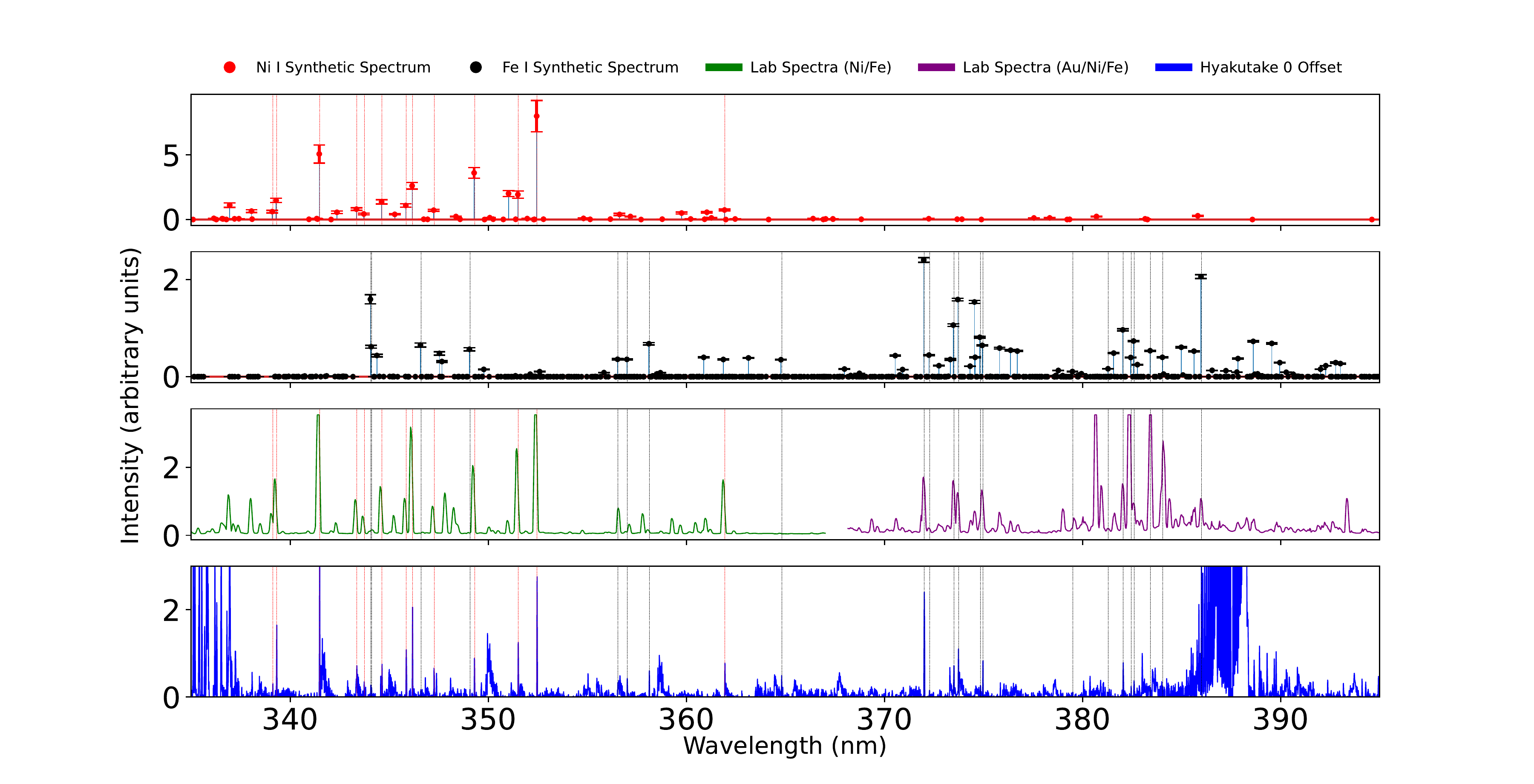}
\caption{{From top to bottom: (1) Ni I synthetic spectrum (red) at the orbital conditions of Hyakutake, (2) Fe I synthetic spectrum (black) at the orbital conditions of Hyakutake, (3) Laboratory spectrum of a Ni/Fe probe at 3 cm depth in the CTH plasma apparatus (shot 18112918, frame 88, green), and laboratory spectrum of a Au/Ni/Fe probe at 6 cm depth in the CTH plasma apparatus (shot 18113013, frame 81, purple), (4) observation spectrum of Hyakutake (blue). Ni I (red) and Fe I (black) lines identified in the comet spectrum are indicated by vertical dashed lines in the lab and comet spectra.}}\label{fig:cth_vs_comet}
\end{figure*}

\subsection{Molecular Contamination}\label{sec:molecules}
NUV and optical cometary spectra consist of reflected sun light superimposed on many emission features of radicals. To determine the reliability of our line identifications, we used the Planetary Spectrum Generator (PSG) to generate synthetic fluorescence spectra of the diatomic molecules OH, CN, C$_2$, NH, and CH~\citep{PSG}. The emission of metals, cations, triatomics (e.g.~C$_3$), or prompt emission of OH are presently {un}available in the PSG. Spectra were generated for production rates set to unity, which were then re-scaled {with respect to OH} by the production rates measured by \cite{Meier1998} and \cite{Schleicher2002}. The {combined} molecular spectra were then manually re-scaled {by a single parameter to} match our comet spectra. This procedure is not intended to accurately replicate the exact features in our comet spectra, but to elucidate sources of contamination in identified metallic lines. The molecular features are well spread across the orders of the echelle and make apparent the uncertainty of the wavelength calibration around some of the metallic features.

To aid future spectroscopy efforts, we have defined the following confidence scale for our line identifications. `A' ratings are stated for isolated lines with no strong nearby features and whose line profile is well fit. `B' ratings are listed for lines with one or more flaws, including possible molecular contamination or blends in the lab spectra. Lastly, `C' ratings are listed for lines with serious molecular contamination, poor fit, or low signal-to-noise. For each letter grade, additional flags are shown in parentheses indicating the source of the molecular contamination e.g. (CN).

%
%
\section{Results}\label{sec:results}
\subsection{Detections of \ion{Ni}{1} and \ion{Fe}{1} lines}\label{results:ni0}

The  metal lines from atomic iron and nickel we identified in the spectrum of {C/1996 B2} (Hyakutake) are listed in Table~\ref{tab:ni0_line_data}. Where possible, we have indicated if a line has previously been observed by \cite{Manfroid2021}, \cite{Guzik2021}, or in the spectra of Ikeya-Seki \citep{Slaughter1969, Preston1967}. Ritz, comet, and lab wavelengths, observed and modeled intensities, upper/lower configuration information, and confidence ratings are also provided.

We identified 14 \ion{Ni}{1} emission lines by comparison with the fluorescence model and lab measurements. Within these 14 lines, 13 were found in both the lab and comet spectra, with the remaining line identified by comparison with our model exclusively. In some cases, deviations between the known air wavelengths of the observed lines and their wavelength in the comet spectrum are $\sim$0.02~nm. These wavelength differences were noted by \cite{Ahearn2015} as resulting from calibration of the numerous orders of the echelle spectrograph. However, the direction and scale of the wavelength shift(s) are obvious when comparing heights and positions of several nearby metal and/or molecular features.

{Fig.~\ref{fig:cth_vs_comet} shows a comparison of our modeled nickel and iron g-factors, laboratory spectra, and comet spectrum. Ni I (red) and Fe I (black) spectra are shown as stick plots, with error bars reported from the Monte-Carlo procedure described previously. Lines confirmed in the comet spectrum are indicated by dashed red (Ni I) or black (Fe I) vertical lines. For the laboratory spectra, two datasets are shown: spectra from erosion of a Ni/Fe probe (green) and that of a Au/Ni/Fe probe (purple). Spectra of the Au/Ni/Fe probe show enhanced Fe emission compared to the pure Ni/Fe plasma, due to rapid erosion of the gold/nickel layers to expose the underlying stainless steel. As the poor broadband wavelength calibration in the spectral range of the Ni/Fe spectrum (green) led to a systematic shift of approximately -0.06~nm blueward for most lines, all $\lambda_\textrm{lab}$ values for Table~\ref{tab:ni0_line_data} were extracted from the Au/Ni/Fe spectra where available.}

A total of 28 \ion{Ni}{1} lines were reported by \cite{Manfroid2021}, 14 of which were observed in {the centered Hyakutake spectrum}. The nine lines of \ion{Ni}{1} observed in 2I/Borisov by \cite{Guzik2021} were all present in the {spectrum} of Hyakutake. We identified {22} \ion{Fe}{1} lines in the spectrum of Hyakutake, 22 of which were also in the line list of \cite{Manfroid2021}. The additional line was also found in Ikeya-Seki, and is among the weakest iron features in {the Hyakutake spectrum.} The non-detection of this line in the sample of \cite{Manfroid2021} is not surprising given Hyakutake's proximity to Earth ($\Delta = 0.11$~AU), large activity ($Q_\textrm{H$_2$O} > 10^{29}$~mol.~s$^{-1}$, ~\citealt{Meier1998}) and favorable heliocentric velocity. We note that every metal line detected in Hyakutake was {also observed in Ikeya-Seki by \cite{Preston1967} and/or \cite{Slaughter1969}.}

{Sixteen (73\%)} of the comet's \ion{Fe}{1} lines were also identified in the lab spectra, though many more \ion{Fe}{1} and \ion{Fe}{2} features {were apparent between 200 - 300~nm in the lab spectra. The electron temperature $T_\textrm{e} > 20$~eV lead to a significant population of \ion{Fe}{2} arising from ionization and excitation of iron eroded from the stainless steel probe underneath the gold and nickel coatings. In the comet spectrum}, \ion{Fe}{1} lines are generally weaker than those of \ion{Ni}{1}, and we find no evidence of \ion{Fe}{2} (cf. Sec.~\ref{subsec:metals}).

\begin{figure}[h]
\centering
\includegraphics[width=0.45\textwidth]{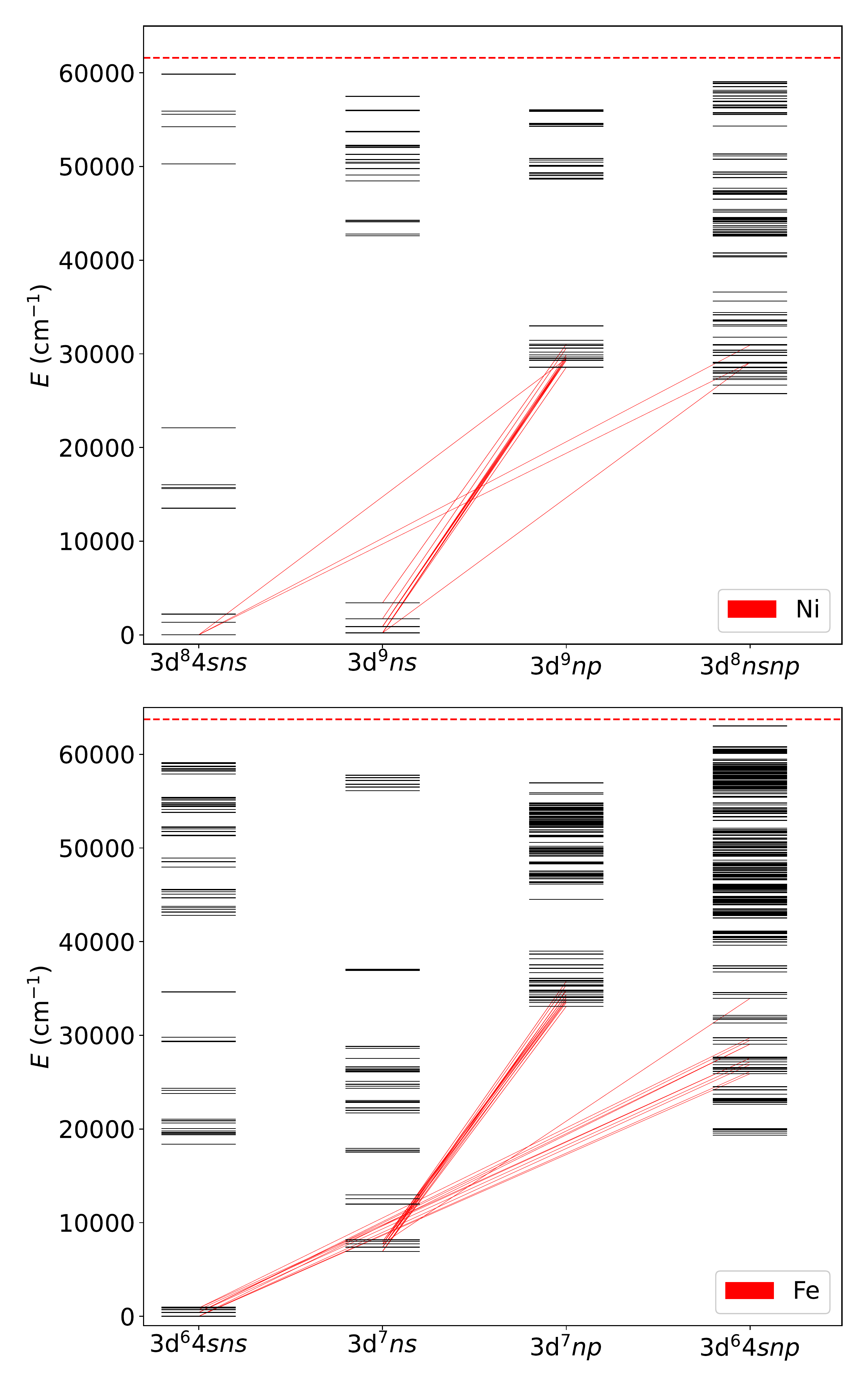}
\caption{Grotrian diagrams of \ion{Ni}{1} (top) and \ion{Fe}{1} (bottom) showing the observed emission lines of each element as diagonal red lines. Ionization limits are shown as horizontal dashed lines. Many lines are known for each element (522 for Ni, 2539 for Fe) but only the observed lines are shown for clarity.}\label{fig:grotrian}
\end{figure}

{The} observed metal emission lines are dipole-allowed transitions from the excited odd configurations of the form $3d^{x}4p$ and $3d^{x}4s4p$ to the ground and metastable configurations $3d^{x}4s$ and $3d^{x}4s^2$. In Fig.~\ref{fig:grotrian} we show Grotrian diagrams of \ion{Ni}{1} and \ion{Fe}{1} where observed transitions are indicated by diagonal red lines. {From our fluorescence models, we find no significant population of highly excited even parity states that would indicate that any observed emission lines from highly-excited even-parity configurations would be produced by non-fluorescence mechanisms.} Similarly, we found no transitions in the comet spectrum which cannot be explained by absorption from the ground or metastable levels; the spectrum is fully consistent with fluorescence of atomic Ni and Fe. The {average agreement between our modeled fluorescence efficiencies and the observed strengths of the cometary metal features agree within a factor of $\sim$2.}

{As electron impact can populate higher excited states compared to fluorescence, the lab spectra exhibit many transitions, e.g. $3d^84s4d \rightarrow 3d^84s4p$ transitions in \ion{Ni}{1}, that are inaccessible via absorption alone. Transitions common to both the lab and comet spectra are typically the strongest transitions in the lab spectra, including the $\lambda = 341.48$~nm and $\lambda = 352.45$~nm lines of the form $3d^94p\rightarrow 3d^94s$. The upper levels of these transitions are accessible by both photon- and electron-impact excitation of the ground configuration.}

According to {the fluorescence model, the first 6 levels of \ion{Ni}{1} above ground have populations comparable to the ground state. For \ion{Fe}{1}, the metastable levels also have sizable populations, but neither \ion{Ni}{1} or \ion{Fe}{1} is predicted to have observable emission in the infrared regime.}

\begin{figure*}[t]
\centering
\includegraphics[width=1\textwidth]{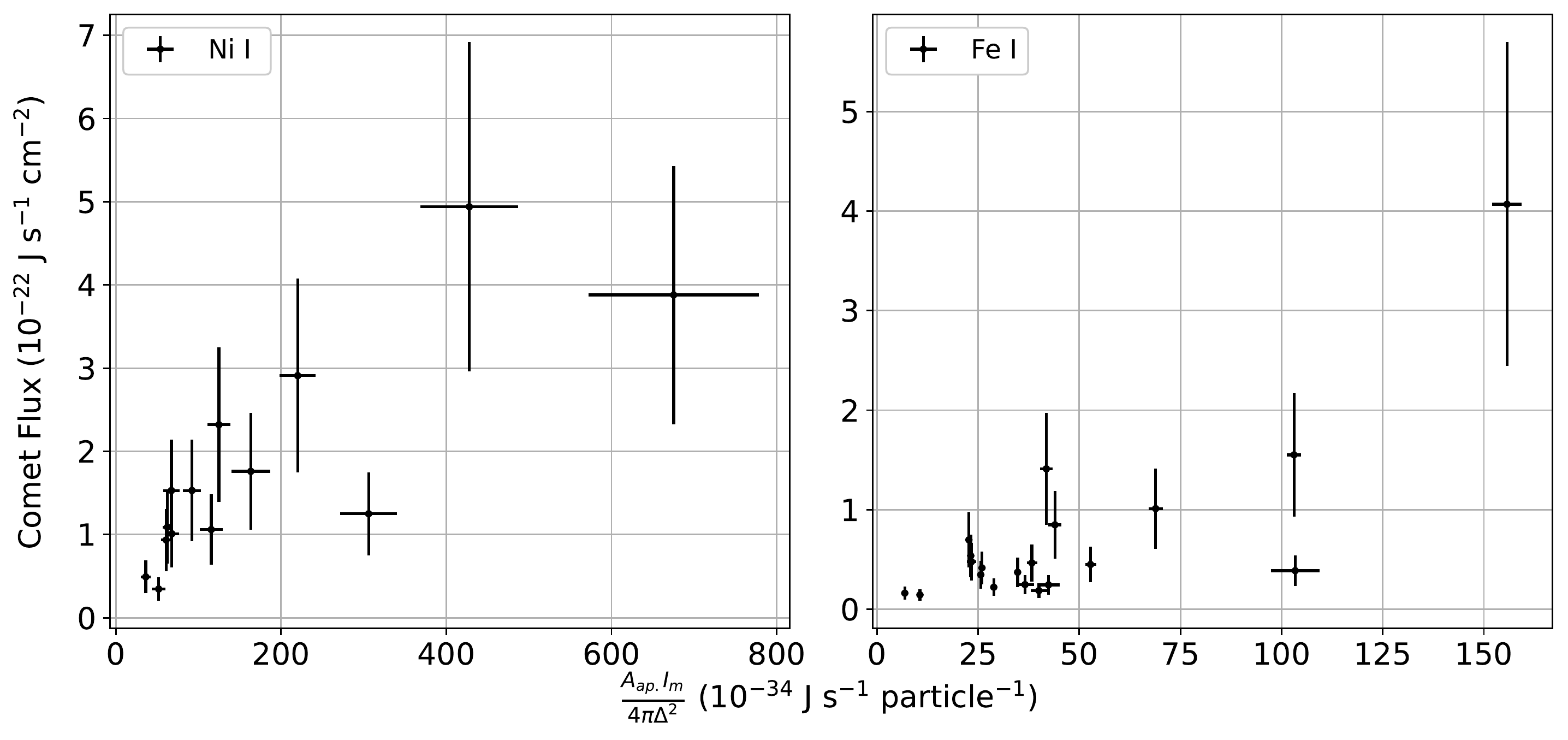}
\caption{Comparison of modeled and observed line intensities of \ion{Ni}{1} (left) and \ion{Fe}{1} (right). Model error bars are taken as the quadrature sum of the Monte-Carlo uncertainty and the sensitivity to heliocentric velocity (see text), and an absolute error for the observed intensities was assumed to be $\pm$40\%~\citep{Meier1998}.}\label{fig:model_vs_comet_comparison}
\end{figure*}

\subsection{Searches for Other Metal Emission}\label{subsec:metals}
{We searched Hyakutake's spectrum} for the emission of  \ion{Fe}{2} and \ion{Ni}{2}; no matches in both wavelength and model intensity were found. We also searched for lines belonging to neutral and singly ionized metals present in Ikeya-Seki and dust grains: Al, Ca, Co, Cr, Cu, Mg, Mn, Na, Zn, and V.

{Searches for \ion{Na}{1} emission were hopeful considering the  substantial $g$-factors for the 589~nm doublet in order 5 -- 10 photons~s$^{-1}$~particle$^{-1}$~\citep{Cremonese1997}, and its detection in  the tail of Hyakutake by \cite{Wyckoff1999}. In comparison, the strongest line of \ion{Ni}{1} in the present spectrum at 341.48~nm has a fluorescence efficiency $g = 3.7\times10^{-21}$~J~s$^{-1}$~particle$^{-1}$= $6\times10^{-3}$~photons~s$^{-1}$~particle$^{-1}$.}

{Other possible \ion{Na}{1} lines in the present wavelength region are driven by emission from autoionizing levels which are unlikely to be populated by absorption alone. Models with numerous subsets of levels were computed, and g-factors for lines around $\sim$390~nm are several to many orders of magnitude weaker than the doublet at 589~nm. Thus, we expect no sodium emission within the 300 -- 400~nm window. For each element/ion above we computed g-factors considering only levels beneath the first ionization potential as higher lying levels are likely to be autoionizing. No matches for wavelength and relative intensity were found.}

We draw attention to a particularly strong and sharp line at 416.68~nm for which no emission lines produced by our models were an appropriate match. Comparisons to data in the ASD suggest \ion{V}{1} or \ion{Cu}{2}, but these two transitions lack $A$ values and both are transitions between two excited states, i.e. they are likely weak at 1~AU. A possible (prompt) molecular origin is NH, though the transition rate for this particular transition is weak~\citep{Fernando2018}.

\subsection{Production Rates and Abundances}\label{subsec:production_rates}
Spectra were taken at various spatial offsets in the Hyakutake observations of \cite{Meier1998}, but there is great uncertainty in photocenter location with respect to the nucleus for each offset observation. We {focused} our analysis on the centered observation and {derived column densities (averaged over the slit) using the relation}

\begin{equation}\label{eq:col_dens}
n_\textrm{col.} = \frac{F_{ij}} {g_{ij}} \frac{4\pi\Delta^2} {A_\textrm{ap.}}
\end{equation}
where $F_i$ is the observed cometary flux from line $\lambda_{ij}$, $A_\textrm{ap.}$ is the sky projected slit size ($68\times580$~km$^{2}$), $\Delta$ is the geocentric distance, {and} $g_{ij}$ is the calculated fluorescence efficiency (Eq.~\ref{eq:intens}). {Column densities, $n_\textrm{col.}$, were calculated from Eq.~\ref{eq:col_dens} as the geometric mean of the column densities derived from all observed metal features. From our mean observed/model line ratios (Fig.~\ref{fig:model_vs_comet_comparison}) and the uncertainties of our calculated g-factors we find $n_\textrm{Ni} = 1.17^{+0.17}_{-0.14}\times10^{10}$~cm$^{-2}$ and $n_\textrm{Fe} = 1.66\pm{0.06}\times10^{10}$~cm$^{-2}$.}

Assuming a gas expansion velocity of $v = 0.85~R_{\textrm{h}}^{-1/2}$~km~s$^{-1}$~\citep{Cochran1993}, we calculate the production rate of these metal atoms using Haser models (see the recent translation in \citealt{Haser_Oset}) implemented in the Small Body Python (\textbf{sbpy}) code~\citep{sbpy}. The use of a Haser model to derive production rates in this context requires several assumptions. For species produced via dissociation or fragmentation (as expected for nickel and iron), each fragment receives a velocity kick (distribution) which may result in a higher outflow velocity than that  derived from the assumed relation above. Presently, the production mechanisms and possible parent loss mechanisms are unknown, and application of a vectorial model~\citep{Festou1981} would not improve the accuracy of our results.

We have attempted to minimize the inaccuracies of our Haser models by restricting our parameter space and deriving daughter scalelengths from known photoionization rates of nickel and iron. {Following a similar procedure to \cite{Guzik2021}, we used} the known lifetimes at 1~AU during solar minimum~\citep{Huebner2015} for \ion{Ni}{1} ($1.06\times10^6$~s) and \ion{Fe}{1}~($5.94\times10^{5}$~s) {to derive} daughter scalelengths of $8.92\times10^5$~km and $5.01\times10^5$~km, respectively. For the unknown parent, we restricted the scalelength to $l_\textrm{parent} < 1000$~km {in line with the procedure reported by \cite{Guzik2021}.}

Without additional spatial information, we are limited to restricting the production rate{s} to a range within the bounds calculated from the uncertainty of the observed total numbers of particles. We iterated the Haser models over a grid of parent scalelength and production rate, after which the Haser column densities were integrated over the rectangular aperture. Using the observed total number of particles within our aperture ($N = A_\textrm{ap.} \times n_\textrm{col.}$, {$N_\textrm{Ni} = 4.6^{+0.7}_{-0.5}\times10^{24}$, $N_\textrm{Fe} = 6.5^{+0.3}_{-0.2}\times10^{24}$}), we narrowed the range of production rates for nickel and iron to $Q_\textrm{Ni} = 2.6 - 4.1\times10^{22}$~s$^{-1}$ and $Q_\textrm{Fe} = {0.4 - 2.8}\times10^{23}$~s$^{-1}$.

\begin{figure*}[t]
\centering
\includegraphics[width=1\textwidth]{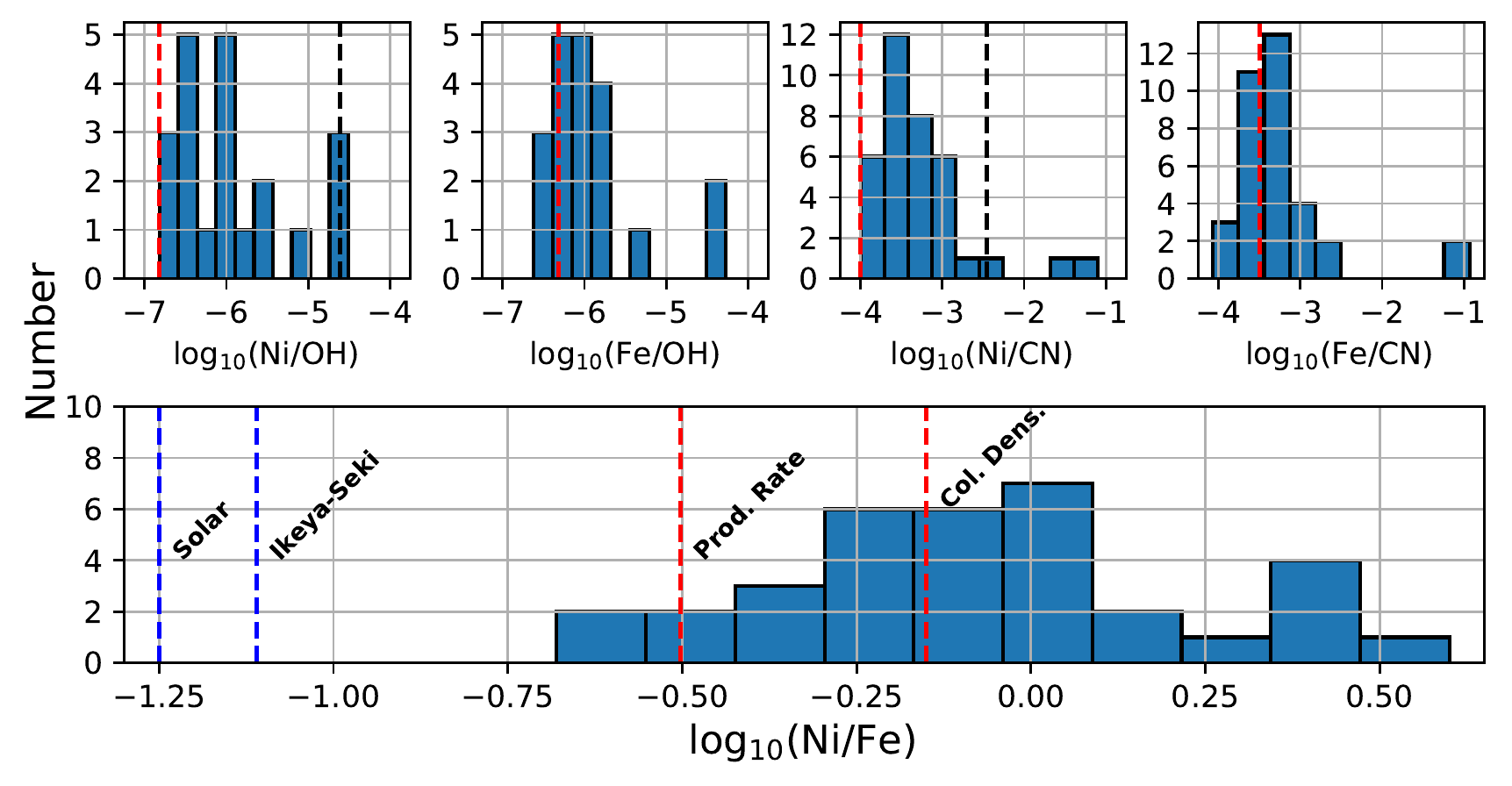}\caption{Histograms (\# bins = 10) of [Ni/OH], [Fe/OH], [Ni/CN], [Fe/CN], and [Ni/Fe] in log$_{10}$ for all comets with measured nickel and iron production rates to-date. Ni, Fe, OH, and CN production rates were taken from \cite{Manfroid2021}, \cite{Meier1998}, \cite{Schleicher2002}, and \cite{Guzik2021}. Ratios for 2I/Borisov are indicated by dashed black vertical line, and ratios for Hyakutake are indicated by red dashed lines. Metal/molecule abundance ratios for Hyakutake were derived from the {geometric mean} of the constrained range of production rate in Hyakutake; for Ni/Fe, ratios derived from the mean production rates and column densities are both shown. }\label{fig:molecule_comp}
\end{figure*}

\section{Discussion}\label{sec:discussion}
\subsection{Comparison of Hyakutake to the broader sample of comets}
Due to their large fluorescence efficiencies, neutral iron and nickel emission were observed for production rates as low as $\sim8\times10^{21}$~s$^{-1}$ by \cite{Manfroid2021}. Our production rates are comparable to those observed in other solar system comets in order $\sim10^{22}$~s$^{-1}$ with the large uncertainties stemming from the lack of spatial information and intensity calibration of {the} echelle spectra. Hyakutake was a particularly active comet with $Q_{\textrm{H$_2$O}} > 2.2\times10^{29}$~mol.~s$^{-1}$~\citep{Meier1998}, { and our mean iron production rate of $Q_\textrm{Fe} = {1.6}\times10^{23}$~s$^{-1}$ is comparable to the most active comet observed by \cite{Manfroid2021}, C/2016 R2 (PANSTARRS). With a mean value of $Q_\textrm{Ni} = 3.4\times10^{22}$~s$^{-1}$,} the nickel production rate in Hyakutake is in-line with the majority of the already-observed solar system comets.

The Ni/Fe abundance ratios reported for solar system comets (in log$_{10}$) spans the range -0.68 to +0.6. From our production rates, the Ni/Fe abundance observed in Hyakutake (in log$_{10}$, hereafter denoted [Ni/Fe]) ranges from {-0.98 to -0.03}. This range is dominated by additional uncertainties introduced via the lack of constraint on parent scalelength in the present work. In this context, it is more appropriate to derive a [Ni/Fe] abundance from the column densities which yields {[Ni/Fe] = $-0.15\pm0.07$.} This Ni/Fe abundance differs significantly from both solar (-1.25) and Ikeya-Seki~(-1.1) ratios where the fluorescing material is sourced from the bulk composition, and from dust collected from 1P/Halley where {log$_{10}$(Ni/Fe) $= -1.1$}~\citep{Jessberger1988}. {Both 0 and 2 arcsecond spectra from the archived observations of Hyakutake were analyzed independently by \cite{Opitom2021_hyak}\footnote{Published during the review process of the present work}, who found Ni/Fe between -0.07 and -0.14 (depending on the offset) using a fluorescence model with restrictions on the input atomic data. These values overlap with the present work within the uncertainties and suggest that the most important transitions were common to both sets of models.}

Fig.~\ref{fig:molecule_comp} shows a comparison of abundance ratios in Hyakutake (dashed red lines) to comets studied by \cite{Manfroid2021} and \cite{Guzik2021}. Abundance ratios for Hyakutake were derived from the {geometric mean of the production rate ranges reported} previously and are shown as dashed vertical red lines. Abundance ratios for 2I/Borisov are shown as dashed black lines where applicable. Hyakutake appears relatively depleted in Ni with respect to both OH and CN. Using our mean Fe production rate ({$1.06\times10^{23}$~s$^{-1}$}), Hyakutake's Fe/OH and Fe/CN abundances are similar to the majority of solar system comets.

Nickel and iron have been observed in comets of all dynamical families at distances between 0.68 -- 3.25~AU. The [Ni/OH] abundance ratio of Hyakutake {(-6.8)} is most similar to the Jupiter Family Comet 103P/Hartley~2 ([Ni/OH] = -6.72) and the dynamically new Oort cloud comet C/2003 K4 (LINEAR) with [Ni/OH] $= -6.68$. Similarly, the [Ni/CN] ratio {(-3.99)} is smaller than most of the comets studied by \cite{Manfroid2021} but is again comparable to 103P (-3.79) and C/2003 K4 (-3.96). The presence in 2I/Borisov implies a common organometallic chemistry during planet formation that may be largely independent of dust and volatile formation chemistry given the differences between most solar system comets and the Ni/OH and Ni/CN abundances of 2I/Borisov. The presence in the CO-rich comet C/2016 R2 (PANSTARRS)~\citep{McKay2019} is puzzling considering the production rates of nickel and iron in C/2016~R2 are among the largest to-date. The high metal production and low water production in C/2016~R2, in addition to some comets having enhanced water production from icy grain sublimation, suggests Ni/OH or Ni/H$_2$O abundance ratios present a poor categorization of the metallic inventory. It is clear that the magnitudes of nickel and iron production are correlated to overall gas and dust production~\citep{Manfroid2021}, but the driving mechanism behind metal atom production in the coma has yet to be confirmed.

\subsection{Chemical origins and physical processes}

Detection of gas-phase nickel and iron vapor around comets presents a new intersection of chemistry, planetary science, and atomic physics. It is well-known that complex molecules form in planetary disks in the gas phase and on grain surfaces~\citep{Berg2017}, and there are indications that transition metal chemistry is gaining interest, e.g. \cite{Boutheina2019} and references therein. Both the parent molecule(s) and the excitation process of the atomic metals are currently unknown. However, it is possible to constrain likely parents and production mechanisms considering the totality of {available laboratory studies.}

\begin{figure}[t]
\centering
\includegraphics[width=0.5\textwidth]{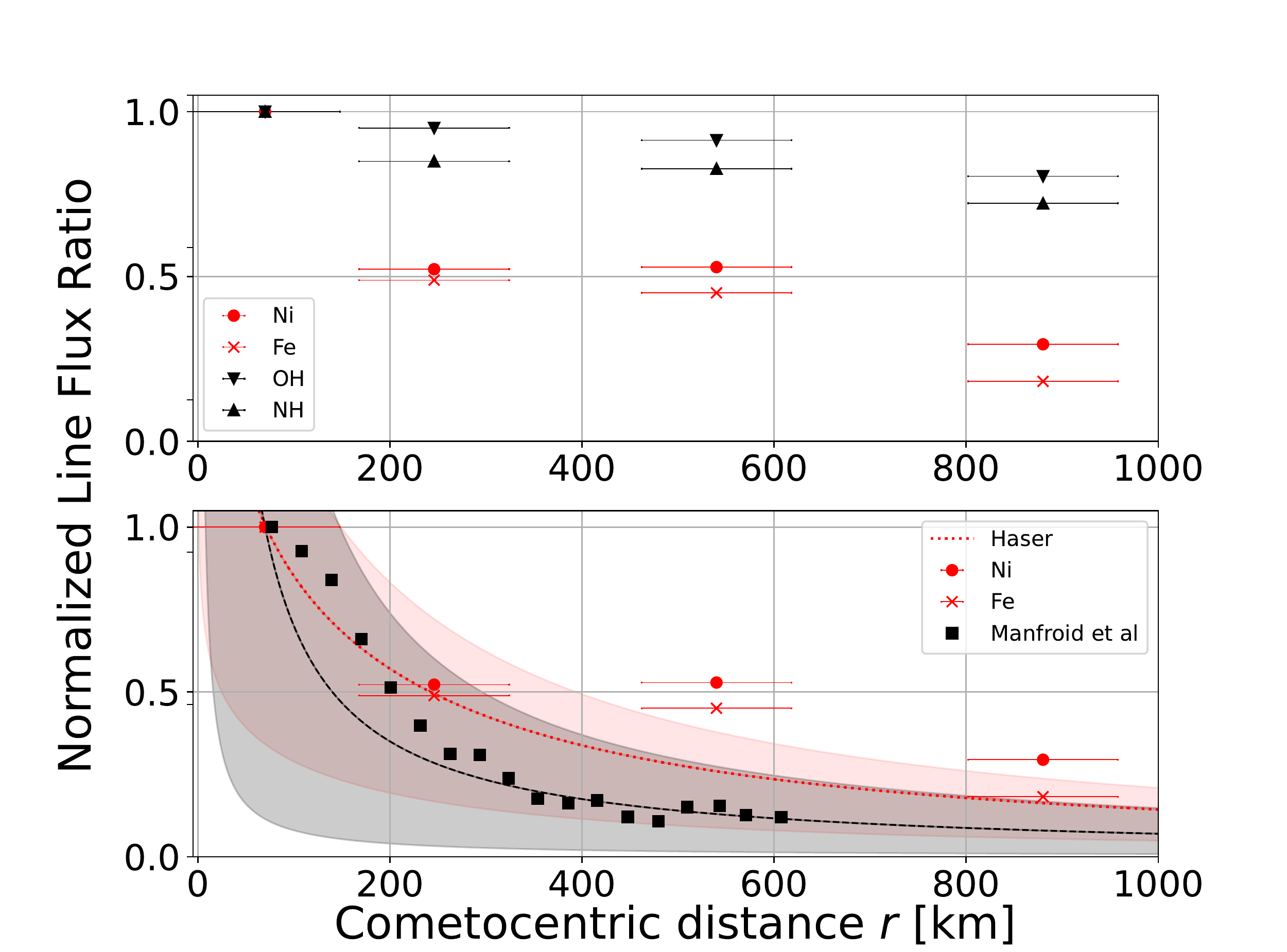}
\caption{$(Top)$ Spatial distributions of nickel, iron, and molecular lines normalized to 1 at photocenter {($r = 70~km)$}. Ni ($\circ$) and Fe ($\times$) flux ratios were taken as the mean of {$I(r)/I(r=70)$ of the 7 Ni and 4 Fe (7) lines present at all 4 distances.} OH ($\triangledown$) and NH ($\triangle$) fluxes and cometocentric distances were digitized from \cite{Meier1998}. $(Bottom)$ Comparison of present Ni and Fe lines {(red $\circ$ and red $\times$, respectively)} to the Fe line profile of comet 103P/Hartley~2 ($\blacksquare$, shifted +70~km) reported in \cite{Manfroid2021}. A $1/r$ profile normalized to 1 at $r = 70$~km is shown in black with grey contours indicating position uncertainty of the 0 arcset pointing of Hyakutake. A Haser model of atomic Ni is shown as a dotted red line, assuming daughter and parent scalelengths of $8.9\times10^5$~km and 200~km with red shaded contours showing the uncertainty of the Ni {Haser} spatial distribution assuming the uncertainty of the photocenter pointing {at 0 arcsec.}}\label{fig:spatial}
\end{figure}

First we consider the spatial distribution of the nickel and iron emission. \cite{Ahearn2015} noted several complications in the exact positions of the pointings: the guiding for the Hyakutake spectra was carried out using visible wavelengths, and it is expected that spatial offsets in the UV are likely $\sim$100~km larger, with variations of order 10s~km within a given spectral range. While tracking the comet for the 7 arcsec offset observation, a fragment was visually identified and noted in the observing log. We therefore expect emission from the 7 arcsec {spectrum} to be enhanced from the presence of a localized transient within the field of view~\citep{Ahearn2015}. It is likely that the true spatial offset differs for each offset observation. Therefore, we use these offset spectra only to comment on a qualitative comparison of the spatial distributions.

Spectra of comet {103P/Hartley~2}, collected at multiple orientations by \cite{Manfroid2021}, indicate that the distribution of nickel and iron is isotropic, and the spatial distribution {follows} a $1/r$ profile indicating either direct sublimation from the nucleus or a short-lived parent. We investigated the presence of our Ni and Fe lines from the {centered observation} in the spectra collected at 2, 7, and 10 arcseconds. Fig.~\ref{fig:spatial} (top) shows the spatial distribution of Ni and Fe lines in the present work compared to OH and NH emission from \cite{Meier1998}. For each species, we normalized the emission to 1 at $I(r=0)$, which is shown at the expected offset distance (70~km) in \cite{Meier1998}. We see large differences between the spatial distributions of Ni and Fe lines compared to the molecular fluxes of OH and NH bands {from} \cite{Meier1998}. Metal features appear comparatively more enhanced than the molecular features at 7 arcseconds from the presence of the local transient.

\cite{Schleicher2002} proposed that sublimation of small icy grains in the coma led to enhancement of the water production in Hyakutake, {and the more recent study of} \cite{Sunshine2021} {elaborates} on the contributions of icy grain sublimation to water vapour around 103P/Hartley~2, which shows an asymmetric H$_2$O profile. The differences between water ice and water vapor {spatial distributions} are clearly distinguishable in imaging from \cite{Ahearn2011, Protopapa2014}. However, the \ion{Fe}{1} profile reported by \cite{Manfroid2021} for 103P was found to be isotropic. If icy grains in the coma contained the Ni and Fe parents, the profiles would be more similar to those observed for e.g. OH and NH. The isotropy of the metal emission and the sharp profiles suggest that the parents of Ni and Fe are not related to icy grain sublimation in the coma.

Fig.~\ref{fig:spatial} (bottom) shows a comparison of the present Ni and Fe flux profiles in Hyakutake and a Fe profile from 103P/Hartley~2 digitized from \cite{Manfroid2021}. A $1/r$ profile (black) and a Haser model (red) are both shown, {with contours showing the effect of pointing uncertainty at 0 arcseconds.} Ignoring the enhanced fluxes at 7 arcsec (540~km) due to the {possible} localized transient, we find reasonable agreement for a simple Haser model with a parent scalelength of order hundreds~km, consistent with the assertion of a short-lived parent {\textit{or} direct release from the nucleus.} Without additional insights into the accuracies of the spatial offsets, a more refined analysis of the spatial distribution in Hyakutake is unreliable.

Two possible parents have been proposed thus far: metal carbonyls, and metal-bonded polycyclic aromatic hydrocarbons (herafter MBPAHs). Iron and nickel carbonyls (Fe(CO)$_5$ and Ni(CO)$_4$) have been suggested as possible parents by \cite{Manfroid2021}, and their chemical model suggests sublimation temperatures in the range 74 -- 108~K, comparable to CO~(81~K), explaining the production of nickel and iron out to several AU. {Carbonyl photochemistry has attracted attention in the laboratory as a target for ultrafast photoionization studies; see e.g. \citep{Distefano1970,Fuss2001,Leadbeater1999} and the most recent study of iron carbonyl by \cite{Cole2021}.  \cite{Kotzian1989} reported absorption cross sections for chromium, iron, and nickel carbonyls (anchored to absolute scale between 224 -- 237~nm depending on species), and investigated the excited populations of removed CO groups.}

{Using the reported absorption cross sections of \cite{Kotzian1989} and assuming their absolute scaling applies across the full 200 -- 350~nm range, we estimate the average number of dissocations per second experienced by a single carbonyl molecule in the presence of solar radiation (at 1~AU). As a caveat, the UV-absorption cross sections of the intermediate complexes are not known. We assume said intermediates have the same UV-absorption characteristics as the initial parent carbonyls and that each UV absorption removes only 1 CO group, though UV absorption can remove as many as 3 or 4 CO groups at once~\citep{Leadbeater1999}.}

{Convolution with our solar spectrum and integration over the 200 -- 350~nm data from \cite{Kotzian1989} yields 0.03 dissociations per second ($\sim$30 s/ionization) for both nickel and iron carbonyls. Assuming a constant velocity of 1~km/s, the initial carbonyl molecules would be fully stripped to bare Ni/Fe within 120 -- 150~km, a distance consistent with the spatial distributions observed thus far. Assuming iron/nickel carbonyls are the precursors to atomic iron/nickel in the coma, successive absorption of UV photons is thus a plausible production mechanism. The similarity in dissocations per second for both metal carbonyls implies that differences in UV absorption characteristics are not the source of the deviation from solar Ni/Fe abundance. Future studies on the intermediate complexes such as Fe(CO)$_4$ would be required to validate the assumptions in the above estimation.}

{We searched for laboratory experiments that would support the possibility of dissociative processes that would result in the direct production of Ni or Fe in excited states, such as electron impact dissociation or photodissociative excitation. \ion{Fe}{1} emission was identified from electron impact dissociative emission of iron carbonyl at 50~eV~\citep{Ribar2015}. Their apparatus operates in the single collision regime where only emission from the immediate dissociation and not electron impact excitation of the fragments is observed before the fragments exit the field of view. \cite{Ribar2015} reported emission lines of \ion{Fe}{1} from emissive dissociation of iron carbonyl, including the strongest \ion{Fe}{1} emission in the spectrum of Hyakutake at $\sim$372~nm. However, other strong transitions are notably absent. The preponderance and variety of \ion{Fe}{1} lines identified to-date in comet spectra, the sensitivity to heliocentric velocity (to be discussed in Sec.~\ref{sec:vel_sens}), and the good agreement with fluorescence models suggest that prompt emission is a small or non-existent contribution to the observed metal emission features.}

{Metal-bonded polycyclic aromatic hydrocarbons (MBPAHs) have been proposed to exist in the interstellar medium (ISM, see e.g.~\citealt{Klotz1996}), and are an alternative proposed precursor to atomic nickel/iron. Transition-metal astrochemistry is an active area of research, and there are indications that metals may play an important role in synthesizing known organic compounds~\citep{Fioroni2014} and depleting metal abundances (in particular, Ni$^+$) in the ISM~\citep{Boutheina2019}.  At present, no possible Ni- or Fe-bearing PAH precursor molecules have been identified, though the abundance of Ni with respect to water is comparable to the abundance of PAHs with respect to water, log$_{10}$[PAH/H$_2$O]$\approx$~-6~\citep{Klotz1996}. One can expect that a MBPAH parent, if truly the precursor of gas-phase metals in comae, would most likely be verified by a (cryogenic) sample return mission, or infrared spectroscopy of the parent informed by quantum chemical structure calculations. Embedded ions have been observed to induce frequency shifts in PAH molecules (see for example experiments on Cu$^+$ in quinoline complexes, \citealt{Gao2016}), but it is unknown if similar differences are observed in nickel/iron-bearing PAHs.}

{Lastly, we investigated the possibility of ionic fragments (Ni$^+$, Fe$^+$) as precursors to the observed neutral metals. For these ions our fluorescence model suggests g-factors many orders of magnitude lower than the neutrals, and ionic fragments would need to neutralize via charge exchange or radiative/dielectronic capture within a distance of order $\sim$hundreds~km to explain the observed spatial profiles. Capture of a free electron to produce neutral metal atoms is unlikely provided the low electron densities in the inner coma, c.f. $n_\textrm{e} \leq 10^3$ at 67P~\citep{Myllys2019}.}

{Ion impact experiments of iron carbonyl~\citep{Indrajith2019} indicate Fe$^+$ energies post-fragmentation between 0.1 -- 1~eV (0.6 -- 1.9~km/s). At these low energies the cross section for charge exchange is likely small, $< 10^{-17}$~cm$^2$ (see \cite{Friedman2017} and references therein). For a nominal neutral density in the inner coma of $10^{10}$~cm$^{-3}$, the mean free path for charge exchange is in order $10^4$~km, too large to explain the observed spatial profiles. Thus we fully expect the atomic iron and nickel are released into the coma as neutral products of photochemistry.}

\begin{figure}
\centering
\includegraphics[width=0.45\textwidth]{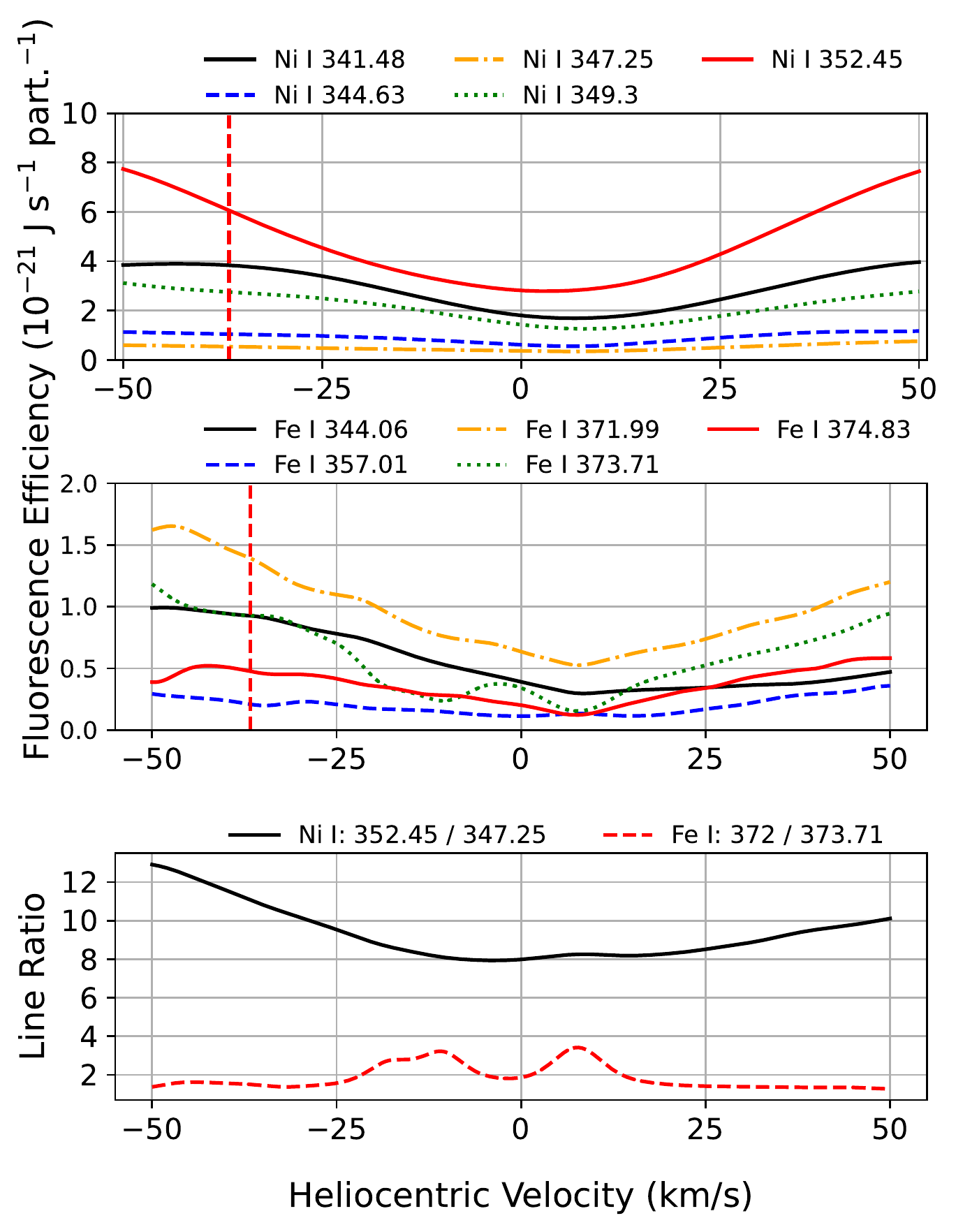}
\caption{Sensitivity of our modeled fluorescence efficiencies for some representative \ion{Ni}{1} lines $(top)$ and \ion{Fe}{1} lines $(middle)$ to heliocentric velocity, and $(bottom)$ example velocity-sensitive line ratios for \ion{Ni}{1} (black) and \ion{Fe}{1} (red). Efficiencies were calculated for a heliocentric distance of 1~AU. The heliocentric velocity of Hyakutake (-36.7~km/s) is shown as a vertical dashed line.}\label{fig:vel_sens}
\end{figure}

\subsection{Sensitivity to Heliocentric Velocity}\label{sec:vel_sens}
{We computed  fluorescence efficiencies of our observed Ni and Fe lines as a function of heliocentric velocity.} Fig.~\ref{fig:vel_sens} shows the calculated fluorescence efficiencies for $r_\textrm{h} = 1$~AU for 5 observed lines of \ion{Ni}{1} (top) and \ion{Fe}{1} (bottom) as a function of heliocentric velocity; calculated fluorescence efficiencies for all of our observed metal lines are available in Appendix~\ref{app:fluor}.

\cite{Manfroid2021} noted that the metal fluorescence efficiencies are sensitive to heliocentric velocity as the driving fluxes for many of the metal emission lines reside at or near absorption features in the solar spectrum. In Fig.~\ref{fig:vel_sens}, as $v_\textrm{h}$ changes the spectral location of the driving flux sweeps over absorption features, {resulting in g-factors that vary by a factor of $\sim$~3 over the range $v_\textrm{h} = \pm50$~km/s}. These changes are comparable to the velocity sensitivity of molecular emission, e.g. a factor of 2 for the NH A-X (0-0) band in the range $v_\textrm{h} = \pm$80~km/s~\citep{Meier1998} and a factor of 4 for OH A-X (0-0) across $v_\textrm{h} = \pm60$~km/s~\citep{Schleicher1988}. {Lines of \ion{Fe}{1} have smaller g-factors but exhibit greater sensitivity to heliocentric velocity, in some cases nearing a factor of 10 (\ion{Fe}{1} 373.71~nm). For both metal atoms, velocity-sensitive g-factors are insensitive to the choice of line profile as either delta function or Doppler-broadened.}

{In light of our previous discussion and the agreement between fluorescence models and observed line ratios, it is likely that the metal emission is purely fluorescent in origin. The observation of two lines, one strongly sensitive to $v_\textrm{h}$ and one insensitive to $v_\textrm{h}$, from a single comet at multiple heliocentric velocities would allow for confirmation of a definitive fluorescence mechanism via comparison with computed g-factors. Two potential line ratios (Fig.~\ref{fig:vel_sens}, bottom panel) for this purpose would be \ion{Ni}{1} $I(352.54)/ I(347.25)$ or \ion{Fe}{1} $I(371.99)/I(373.71)$. Any line ratio involving \ion{Fe}{1} 371.99~nm is favorable as the line is among the strongest observed iron features (cf. the line list of \citealt{Manfroid2021}\footnote{Line list available as supplementary information at \url{https://www.researchsquare.com/article/rs-101492/v1}}) and the g-factor exhibits a strong dependence on heliocentric velocity.
}


\section{Summary}\label{sec:summary}
We developed a many-level fluorescence model compatible with atomic data in the NIST ASD~\citep{NIST_ASD} {and made the code publicly available.} Using our fluorescence model, we {searched archived data of the 1996 apparition of comet Hyakutake for possible metal lines identified in other comets by \cite{Manfroid2021} and \cite{Guzik2021}.} The nucleus-centered {spectrum} of C/1996 B2 (Hyakutake) {was} compared against synthetic fluorescence spectra of atomic metals {and spectra from laboratory plasmas seeded with nickel and iron}. We identified 14 emission lines of \ion{Ni}{1} and {22} lines of \ion{Fe}{1} {in the comet spectrum}. Agreement between {fluorescence models and observed features of \ion{Ni}{1} and \ion{Fe}{1} was} achieved within a factor of $\sim$2 on average.

Assuming fluorescence emission, we derived column densities of nickel and iron atoms, from which we used the total number of particles within our aperture to inform Haser models to estimate the production rates {$Q_\textrm{Ni} = 2.6 - 4.1\times10^{22}$~s$^{-1}$ and $Q_\textrm{Fe} = 0.4 - 2.8\times10^{23}$~s$^{-1}$}. Using our derived column densities {to determine} the Ni/Fe abundance ratio, {we find log$_{10}$[Ni/Fe] = $-0.15\pm0.07$},  a value that differs significantly from abundances observed for the sun, C/1965 S1 (Ikeya-Seki), and in-situ measurements of {dust in} 1P/Halley. Our {abundance} ratios are in-line with those observed in other solar system comets and 2I/Borisov.

{We considered the possible sources and excitation mechanisms of the atomic iron and nickel. The observed spectrum is fully consistent with photofluorescence, with no evidence for the presence of highly excited states that could indicate a dissociative excitation process. We suggest that the strong dependence of the fluorescence efficiency on the heliocentric velocity can provide a direct test of this mechanism.}

{The distribution of the emission of \ion{Fe}{1} and \ion{Ni}{1} can be explained by the dissociation of a short-lived parent. Two possible parents have been proposed thus far: metal carbonyls, and PAH complexes bearing metal atoms. PAH abundances with respect to water are of similar magnitude to Ni/Fe \citep{Bodewits2021,Klotz1996}, and remain a possible precursor to atomic metals in the coma. Metal carbonyls, alternatively, are expected to sublimate at temperatures similar to CO ice and have the requisite characteristics to explain the observed Ni/Fe abundances. UV absorption cross sections of iron and nickel carbonyl suggest that these precursor molecules, if present, are fully stripped of CO groups to produce atomic nickel and iron within 120 -- 150~km. UV absorption cross sections of the intermediate dissociation products are required to support or refute this possibility. The similarity in UV absorption cross sections for nickel/iron carbonyls suggest the depletion of Ni/Fe with respect to solar is not driven by differences in the absorption properties but is imprinted during initial formation of the precursors.}

The discoveries of \ion{Ni}{1} and \ion{Fe}{1} emission in comets have elucidated a new diagnostic to cometary scientists in which some information on the organo-metallic inventories of comets can be probed without waiting for rare sungrazer events or sample return missions. Given the propensity of nickel and iron emission in the UV-VIS range, these lines may be analyzed alongside typical molecular features. With sufficient sampling, these discoveries may shed light on aspects of as-yet-unexplored organic astrochemistry.

\begin{longrotatetable}
\begin{deluxetable*}{lllllllllllll}
\centering
\footnotesize
\tablecaption{List of Observed \ion{Ni}{1} and \ion{Fe}{1} Lines. All wavelengths are reported as nm in standard air. Einstein $A$ values are written as $m+n \equiv m\times10^n$. Lines previously observed in solar system comets~(SS, \citealt{Manfroid2021}), 2I/Borisov~(BOR, \citealt{Guzik2021}), or Ikeya-Seki~(IS, \citealt{Preston1967,Slaughter1969}) are shown by a `\checkmark'  symbol. Unknown LS term identifications for a given J-level are indicated by a ? symbol.}\label{tab:ni0_line_data}.
\tablehead{\colhead{Species} & \colhead{$^*\lambda_{\textrm{Ritz}}$} & \colhead{{$\lambda_{\textrm{comet}}$}} & \colhead{$\lambda_{\textrm{lab}}$} & \colhead{${^\dag}I_{\textrm{comet}}$} & \colhead{${^\ddag}g_{\textrm{model}}$} & \colhead{$A$ Value} &  \colhead{Lower Level} & \colhead{Upper Level} & \colhead{Confidence} & \colhead{SS} & \colhead{BOR} & \colhead{IS}}
\\
\startdata
\ion{Fe}{1}&344.06&344.07&344.15&38.6$\pm$15.44&0.89$\pm$0.05&1.7+07&$3d^64s^2 ~a~^5D_4$&$3d^6(^5D)4s4p(^3P) ~w~^5P_3$&A (Bl. Lab)&\checkmark&&\checkmark\\ \
\ion{Fe}{1}&344.1&344.11&344.15&18.5$\pm$7.4&0.35$\pm$0.02&1.2+07&$3d^64s^2 ~a~^5D_3$&$3d^6(^5D)4s4p(^3P) ~w~^5P_2$&A (Bl. Lab)&\checkmark&&\checkmark\\ \
\ion{Fe}{1}&346.59&346.6&346.6&24.3$\pm$9.72&0.37$\pm$0.02&1.2+07&$3d^64s^2 ~a~^5D_1$&$3d^6(^5D)4s4p(^3P) ~z~^5P_1$&B (OH; Bl. Lab)&\checkmark&&\checkmark\\ \
\ion{Fe}{1}&349.06&349.07&&24.6$\pm$9.84&0.32$\pm$0.02&6.1+06&$3d^64s^2 ~a~^5D_3$&$3d^6(^5D)4s4p(^3P) ~z~^5P_3$&A&\checkmark&&\checkmark\\ \
\ion{Fe}{1}&356.54&356.55&356.51&47.6$\pm$19.04&0.2$\pm$0.01&4.3+07&$3d^7(^4F)4s ~a~^5F_3$&$3d^7(^4F)4p ~z~^3G_4$&C&\checkmark&&\checkmark\\ \
\ion{Fe}{1}&357.01&357.02&&53.6$\pm$21.44&0.2$\pm$0.01&6.8+07&$3d^7(^4F)4s ~a~^5F_4$&$3d^7(^4F)4p ~w~^3G_5$&A&\checkmark&&\checkmark\\ \
\ion{Fe}{1}&358.12&358.13&&84.7$\pm$33.88&0.38$\pm$0.01&1.0+08&$3d^7(^4F)4s ~a~^5F_5$&$3d^7(^4F)4p ~w~^5G_6$&B (CN)&\checkmark&&\checkmark\\ \
\ion{Fe}{1}&364.78&364.81&&69.6$\pm$27.84&0.2$\pm$0.01&2.9+07&$3d^7(^4F)4s ~a~^5F_4$&$3d^7(^4F)4p ~z~^5G_5$&C (CH)&\checkmark&&\checkmark\\ \
\ion{Fe}{1}&371.99&372.01&372.0&407.0$\pm$162.8&1.34$\pm$0.03&1.6+07&$3d^64s^2 ~a~^5D_4$&$3d^6(^5D)4s4p(^3P) ~w~^5F_5$&B (NH)&\checkmark&&\checkmark\\ \
\ion{Fe}{1}&372.26&372.28&&22.0$\pm$8.8&0.25$\pm$0.01&5.0+06&$3d^64s^2 ~a~^5D_2$&$3d^6(^5D)4s4p(^3P) ~z~^5F_2$&C&\checkmark&&\checkmark\\ \
\ion{Fe}{1}&373.49&373.51&373.49&101.0$\pm$40.4&0.59$\pm$0.01&9.0+07&$3d^7(^4F)4s ~a~^5F_5$&$3d^7(^4F)4p ~y~^5F_5$&B (NH)&\checkmark&&\checkmark\\ \
\ion{Fe}{1}&373.71&373.74&373.7&155.0$\pm$62.0&0.89$\pm$0.02&1.4+07&$3d^64s^2 ~a~^5D_3$&$3d^6(^5D)4s4p(^3P) ~w~^5F_4$&A&\checkmark&&\checkmark\\ \
\ion{Fe}{1}&374.83&374.85&&44.8$\pm$17.92&0.46$\pm$0.01&9.2+06&$3d^64s^2 ~a~^5D_1$&$3d^6(^5D)4s4p(^3P) ~w~^5F_2$&C (NH)&\checkmark&&\checkmark\\ \
\ion{Fe}{1}&374.95&374.98&374.95&141.0$\pm$56.4&0.36$\pm$0.01&7.6+07&$3d^7(^4F)4s ~a~^5F_4$&$3d^7(^4F)4p~y~^5F_4$&B (NH)&\checkmark&&\checkmark\\ \
\ion{Fe}{1}&379.5&379.51&&16.0$\pm$6.4&0.06$\pm$0.0&1.2+07&$3d^7(^4F)4s ~a~^5F_2$&$3d^7(^4F)4p~y~^5F_3$&C&\checkmark&&\checkmark\\ \
\ion{Fe}{1}&381.3&381.3&381.27&14.3$\pm$5.72&0.09$\pm$0.0&7.9+06&$3d^7(^4F)4s ~a~^5F_3$&$3d^6(^5D)4s4p(^3P) ~z~^3P_2$&C (Weak in Lab)&&&\checkmark\\ \
\ion{Fe}{1}&382.04&382.06&382.05&111.0$\pm$44.4&0.54$\pm$0.01&6.8+07&$3d^7(^4F)4s ~a~^5F_5$&$3d^7(^4F)4p~y~^5D_4$&A&\checkmark&&\checkmark\\ \
\ion{Fe}{1}&382.44&382.45&&34.5$\pm$13.8&0.22$\pm$0.01&2.8+06&$3d^64s^2 ~a~^5D_4$&$3d^6(^5D)4s4p(^3P) ~z~^5D_3$&A&\checkmark&&\checkmark\\ \
\ion{Fe}{1}&382.59&382.6&382.58&63.3$\pm$25.32&0.41$\pm$0.01&6.0+07&$3d^7(^4F)4s ~a~^5F_4$&$3d^7(^4F)4p~y~^5D_3$&A&\checkmark&&\checkmark\\ \
\ion{Fe}{1}&383.42&383.43&383.42&37.0$\pm$14.8&0.3$\pm$0.01&4.5+07&$3d^7(^4F)4s ~a~^5F_3$&$3d^7(^4F)4p~y~^5D_2$&A&\checkmark&&\checkmark\\ \
\ion{Fe}{1}&384.04&384.06&384.08&41.4$\pm$16.56&0.22$\pm$0.01&4.7+07&$3d^7(^4F)4s ~a~^5F_2$&$3d^7(^4F)4p~y~^5D_1$&A&\checkmark&&\checkmark\\ \
\ion{Fe}{1}&385.99&386.0&385.99&502.0$\pm$200.8&1.15$\pm$0.02&9.7+06&$3d^64s^2 ~a~^5D_4$&$3d^6(^5D)4s4p(^3P) ~z~^5D_4$&C (CN)&\checkmark&&\checkmark\\ \
\ion{Ni}{1}&323.29&323.31&323.35&153.0$\pm$61.2&0.58$\pm$0.08&7.3+06&$3d^8(^3F)4s^2 (^3F_4)$&$3d^8(^3F)4s4p(^3P) (^3G_5)$&A&\checkmark&&\checkmark\\ \
\ion{Ni}{1}&339.1&339.13&339.09&34.6$\pm$13.84&0.45$\pm$0.07&6.6+06&$3d^8(^3F)4s^2 (^3F_4)$&$3d^9(^2D)4p~^3F_4$&B (NH)&\checkmark&&\checkmark\\ \
\ion{Ni}{1}&339.3&339.32&339.27&232.0$\pm$92.8&1.08$\pm$0.12&2.4+07&$3d^9(^2D)4s (^3D_3)$&$3d^9(^2D)4p (\textrm{?})_3$&B (NH)&\checkmark&\checkmark&\checkmark\\ \
\ion{Ni}{1}&341.48&341.49&341.46&494.0$\pm$197.6&3.69$\pm$0.51&5.5+07&$3d^9(^2D)4s (^3D_3)$&$3d^9(^2D)4p (^3F_4)$&A&\checkmark&\checkmark&\checkmark\\ \
\ion{Ni}{1}&343.36&343.37&343.33&101.0$\pm$40.4&0.59$\pm$0.07&1.7+07&$3d^9(^2D)4s (^3D_3)$&$3d^9(^2D)4p (^3F_3)$&A&\checkmark&&\checkmark\\ \
\ion{Ni}{1}&343.73&343.74&343.71&49.2$\pm$19.68&0.31$\pm$0.05&4.4+06&$3d^8(^3F)4s^2 (^3F_4)$&$3d^8(^3F)4s4p(^3P) (^5F_4)$&C&\checkmark&&\checkmark\\ \
\ion{Ni}{1}&344.63&344.64&344.61&106.0$\pm$42.4&1.0$\pm$0.12&4.4+07&$3d^9(^2D)4s (^3D_2)$&$3d^9(^2D)4p (^3D_2)$&A&\checkmark&\checkmark&\checkmark\\ \
\ion{Ni}{1}&345.85&345.86&345.82&153.0$\pm$61.2&0.8$\pm$0.1&6.1+07&$3d^9(^2D)4s (^3D_1)$&$3d^9(^2D)4p (^3F_2)$&A&\checkmark&\checkmark&\checkmark\\ \
\ion{Ni}{1}&346.17&346.18&346.14&291.0$\pm$116.4&1.9$\pm$0.19&2.7+07&$3d^9(^2D)4s (^3D_3)$&$3d^8(^3F)4s4p(^3P) (^5F_4)$&A&\checkmark&\checkmark&\checkmark\\ \
\ion{Ni}{1}&347.25&347.27&347.23&93.5$\pm$37.4&0.53$\pm$0.05&1.2+07&$3d^9(^2D)4s (^3D_2)$&$3d^9(^2D)4p (?)_3$&A&\checkmark&&\checkmark\\ \
\ion{Ni}{1}&349.3&349.31&&125.0$\pm$50.0&2.64$\pm$0.3&9.8+07&$3d^9(^2D)4s (^3D_2)$&$3d^9(^2D)4p (^3P_1)$&C&\checkmark&\checkmark&\checkmark\\ \
\ion{Ni}{1}&351.51&351.52&351.48&176.0$\pm$70.4&1.41$\pm$0.2&4.2+07&$3d^9(^2D)4s (^3D_2)$&$3d^9(^2D)4p (^3F_3)$&A&\checkmark&\checkmark&\checkmark\\ \
\ion{Ni}{1}&352.45&352.47&352.43&388.0$\pm$155.2&5.83$\pm$0.89&1.0+08&$3d^9(^2D)4s (^3D_3)$&$3d^9(^2D)4p (^3P_2)$&A&\checkmark&\checkmark&\checkmark\\ \
\ion{Ni}{1}&361.94&361.95&361.9&109.0$\pm$43.6&0.54$\pm$0.05&6.6+07&$3d^9(^2D)4s (^1D_2)$&$3d^9(^2D)4p (^1F_3)$&A&\checkmark&\checkmark&\checkmark\\ \

\\
\enddata
\tablecomments{\\ * - $\lambda_\textrm{Ritz}$ are calculated (in vacuum) from level energies and converted to air wavelengths using the formulas in \cite{Morton2000}. \\ $\dag$ - $I_\textrm{comet}$ reported in units of 10$^{-24}$~J~s$^{-1}$~cm$^{-2}$ after integration over a gaussian profile fit to each line. \\ $\ddag$ - g-factors are reported in units of $10^{-21}$~J~s$^{-1}$~particle$^{-1}$; uncertainties are taken as the quadrature sum of heliocentric velocity sensitivity (see Sec.~\ref{subsec:total_uncert}) and standard deviation of 10$^6$ Monte-Carlo iterations (see Sec.~\ref{sec:error}).
}

\end{deluxetable*}
\end{longrotatetable}


\software{Python3, NumPy~\citep{numpy}, SciPy~\citep{scipy}, Small Body Python (sbpy, \citealt{sbpy})}

\acknowledgments
We thank John Noonan (Lunar and Planetary Lab, U. Arizona) for his helpful discussions during the development of the fluorescence model. Special thanks are extended to Neil Cole-Filipiak and Krupa Ramasesha of Sandia National Laboratories for engaging discussions of iron/nickel carbonyl photochemistry, and Damien Hutsemekers for thoughtful comments on the fluorescence modeling. The authors gratefully acknowledge funding support from the National Science Foundation (grant Nos. 1815833, 1815932, and
1816984. The Compact Toroidal Hybrid measurements were supported by U.S. DOE Grant No. DE-FG-02-00ER54610. We are grateful for the support of the Auburn University Hopper Cluster for their assistance and computing resources utilized for this work. We extend a special thank you to the observing team~\citep{AhearnPDS2015} for collecting and archiving their data through the Planetary Data System, and the NIST ASD team for acquiring, compiling, and archiving useful atomic data.

\bibliographystyle{aasjournal}
\bibliography{ni_bib}

\begin{thebibliography}{}
\expandafter\ifx\csname natexlab\endcsname\relax\def\natexlab#1{#1}\fi
\providecommand{\url}[1]{\href{#1}{#1}}
\providecommand{\dodoi}[1]{doi:~\href{http://doi.org/#1}{\nolinkurl{#1}}}
\providecommand{\doeprint}[1]{\href{http://ascl.net/#1}{\nolinkurl{http://ascl.net/#1}}}
\providecommand{\doarXiv}[1]{\href{https://arxiv.org/abs/#1}{\nolinkurl{https://arxiv.org/abs/#1}}}

\bibitem[{A'Hearn {et~al.}(1995)A'Hearn, Millis, Schleicher, Osip, \&
  Birch}]{Ahearn1995}
A'Hearn, M.~F., Millis, R.~C., Schleicher, D.~G., Osip, D.~J., \& Birch, P.~V.
  1995, Icarus, 118, 223, \dodoi{https://doi.org/10.1006/icar.1995.1190}

\bibitem[{A'Hearn {et~al.}(2015{\natexlab{a}})A'Hearn, Swamy, Wellnitz, \&
  Meier}]{Ahearn2015}
A'Hearn, M.~F., Swamy, K. S.~K., Wellnitz, D.~D., \& Meier, R.
  2015{\natexlab{a}}, The Astronomical Journal, 150, 5

\bibitem[{A'Hearn {et~al.}(2013)A'Hearn, Wellnitz, \& Meier}]{AHearn2013}
A'Hearn, M.~F., Wellnitz, D.~D., \& Meier, R. 2013, Proceedings of the
  International Astronomical Union, 9, 216–218,
  \dodoi{10.1017/S1743921313015883}

\bibitem[{A'Hearn {et~al.}(2015{\natexlab{b}})A'Hearn, Wellnitz, \&
  Meier}]{AhearnPDS2015}
---. 2015{\natexlab{b}}, NASA Planetary Data System,
  urn:nasa:pds:gbo-kpno:hyakutake\_spectra::1.0

\bibitem[{A{\textquoteright}Hearn {et~al.}(2011)A{\textquoteright}Hearn,
  Belton, Delamere, Feaga, Hampton, Kissel, Klaasen, McFadden, Meech, Melosh,
  Schultz, Sunshine, Thomas, Veverka, Wellnitz, Yeomans, Besse, Bodewits,
  Bowling, Carcich, Collins, Farnham, Groussin, Hermalyn, Kelley, Kelley, Li,
  Lindler, Lisse, McLaughlin, Merlin, Protopapa, Richardson, \&
  Williams}]{Ahearn2011}
A{\textquoteright}Hearn, M.~F., Belton, M. J.~S., Delamere, W.~A., {et~al.}
  2011, Science, 332, 1396, \dodoi{10.1126/science.1204054}

\bibitem[{Altwegg {et~al.}(2019)Altwegg, Balsiger, \& Fuselier}]{Altwegg2019}
Altwegg, K., Balsiger, H., \& Fuselier, S.~A. 2019, Annual Review of Astronomy
  and Astrophysics, 57, 113, \dodoi{10.1146/annurev-astro-091918-104409}

\bibitem[{Berger {et~al.}(2011)Berger, Zega, Keller, \& Lauretta}]{Berger2011}
Berger, E.~L., Zega, T.~J., Keller, L.~P., \& Lauretta, D.~S. 2011, Geochimica
  et Cosmochimica Acta, 75, 3501 ,
  \dodoi{https://doi.org/10.1016/j.gca.2011.03.026}

\bibitem[{Bockelée-Morvan \& Biver(2017)}]{Bockelee2017}
Bockelée-Morvan, D., \& Biver, N. 2017, Philosophical Transactions of the
  Royal Society A: Mathematical, Physical and Engineering Sciences, 375,
  20160252, \dodoi{10.1098/rsta.2016.0252}

\bibitem[{{Bodewits} \& {Bromley}(2021)}]{Bodewits2021}
{Bodewits}, D., \& {Bromley}, S.~J. 2021, \nat, 593, 349,
  \dodoi{10.1038/d41586-021-01265-8}

\bibitem[{Bodewits {et~al.}(2019)Bodewits, Orszagh, Noonan, \v{D}urian, \&
  Matej\v{c}ík}]{Bodewits2019}
Bodewits, D., Orszagh, J., Noonan, J., \v{D}urian, M., \& Matej\v{c}ík, v.
  2019, Astrophysical Journal, 885, 167, \dodoi{10.3847/1538-4357/ab43c9}

\bibitem[{Bromley {et~al.}(2020)Bromley, Johnson, Ennis, Hartwell, Maurer,
  Loch, Stancil, McLaughlin, Sosolik, \& Marler}]{Bromley2020}
Bromley, S.~J., Johnson, C.~A., Ennis, D.~A., {et~al.} 2020, The Astrophysical
  Journal Supplement Series, 250, 19, \dodoi{10.3847/1538-4365/abaa4d}

\bibitem[{Brownlee(2014)}]{Brownlee2014}
Brownlee, D. 2014, Annual Review of Earth and Planetary Sciences, 42, 179,
  \dodoi{10.1146/annurev-earth-050212-124203}

\bibitem[{Chance \& Kurucz(2010)}]{Chance2010}
Chance, K., \& Kurucz, R. 2010, Journal of Quantitative Spectroscopy and
  Radiative Transfer, 111, 1289,
  \dodoi{https://doi.org/10.1016/j.jqsrt.2010.01.036}

\bibitem[{{Cochran} \& {Schleicher}(1993)}]{Cochran1993}
{Cochran}, A.~L., \& {Schleicher}, D.~G. 1993, \icarus, 105, 235,
  \dodoi{10.1006/icar.1993.1121}

\bibitem[{Coddington {et~al.}(2021)Coddington, Richard, Harber, Pilewskie,
  Woods, Chance, Liu, \& Sun}]{Coddington2021}
Coddington, O.~M., Richard, E.~C., Harber, D., {et~al.} 2021, Geophysical
  Research Letters, 48, e2020GL091709,
  \dodoi{https://doi.org/10.1029/2020GL091709}

\bibitem[{Cole-Filipiak {et~al.}(2021)Cole-Filipiak, Troß, Schrader, McCaslin,
  \& Ramasesha}]{Cole2021}
Cole-Filipiak, N.~C., Troß, J., Schrader, P., McCaslin, L.~M., \& Ramasesha,
  K. 2021, The Journal of Chemical Physics, 154, 134308,
  \dodoi{10.1063/5.0041074}

\bibitem[{Cremonese {et~al.}(2002)Cremonese, Huebner, Rauer, \&
  Boice}]{Cremonese2002}
Cremonese, G., Huebner, W., Rauer, H., \& Boice, D. 2002, Advances in Space
  Research, 29, 1187, \dodoi{https://doi.org/10.1016/S0273-1177(02)00136-9}

\bibitem[{Cremonese {et~al.}(1997)Cremonese, Boehnhardt, Crovisier, Rauer,
  Fitzsimmons, Fulle, Licandro, Pollacco, Tozzi, \& West}]{Cremonese1997}
Cremonese, G., Boehnhardt, H., Crovisier, J., {et~al.} 1997, The Astrophysical
  Journal, 490, L199, \dodoi{10.1086/311040}

\bibitem[{Crismani {et~al.}(2018)Crismani, Schneider, Evans, Plane,
  Carrillo‐Sánchez, Jain, Deighan, \& Yelle}]{Crismani2018}
Crismani, M. M.~J., Schneider, N.~M., Evans, J.~S., {et~al.} 2018, Journal of
  Geophysical Research: Planets, 123, 2613, \dodoi{10.1029/2018je005750}

\bibitem[{Distefano(1970)}]{Distefano1970}
Distefano, G. 1970, Journal of research of the National Bureau of Standards.
  Section A, Physics and chemistry, 74A, 233, \dodoi{10.6028/jres.074A.019}

\bibitem[{{Feldman} {et~al.}(2004){Feldman}, {Cochran}, \&
  {Combi}}]{Feldman2004}
{Feldman}, P.~D., {Cochran}, A.~L., \& {Combi}, M.~R. 2004, {Spectroscopic
  investigations of fragment species in the coma}, ed. M.~C. {Festou}, H.~U.
  {Keller}, \& H.~A. {Weaver}, 425

\bibitem[{Fernando {et~al.}(2018)Fernando, Bernath, Hodges, \&
  Masseron}]{Fernando2018}
Fernando, A.~M., Bernath, P.~F., Hodges, J.~N., \& Masseron, T. 2018, Journal
  of Quantitative Spectroscopy and Radiative Transfer, 217, 29,
  \dodoi{https://doi.org/10.1016/j.jqsrt.2018.05.021}

\bibitem[{{Festou}(1981)}]{Festou1981}
{Festou}, M.~C. 1981, \aap, 95, 69

\bibitem[{Fioroni(2014)}]{Fioroni2014}
Fioroni, M. 2014, Phys. Chem. Chem. Phys., 16, 24312,
  \dodoi{10.1039/C4CP03218G}

\bibitem[{Fontenla {et~al.}(2014)Fontenla, Landi, Snow, \&
  Woods}]{Fontenla2014}
Fontenla, J.~M., Landi, E., Snow, M., \& Woods, T. 2014, Solar Physics, 289,
  515, \dodoi{10.1007/s11207-013-0431-4}

\bibitem[{Friedman \& DuCharme(2017)}]{Friedman2017}
Friedman, B., \& DuCharme, G. 2017, Journal of Physics B: Atomic, Molecular and
  Optical Physics, 50, 115202, \dodoi{10.1088/1361-6455/aa6cce}

\bibitem[{Fuss {et~al.}(2001)Fuss, Schmid, \& Trushin}]{Fuss2001}
Fuss, W., Schmid, W.~E., \& Trushin, S.~A. 2001, The Journal of Physical
  Chemistry A, 105, 333, \dodoi{10.1021/jp002276z}

\bibitem[{Gao {et~al.}(2016)Gao, Bouwman, Berden, \& Oomens}]{Gao2016}
Gao, J., Bouwman, J., Berden, G., \& Oomens, J. 2016, The Journal of Physical
  Chemistry A, 120, 7800, \dodoi{10.1021/acs.jpca.6b05060}

\bibitem[{Guzik \& Drahus(2021)}]{Guzik2021}
Guzik, P., \& Drahus, M. 2021, Nature, 593, 375,
  \dodoi{10.1038/s41586-021-03485-4}

\bibitem[{Hall \& Anderson(1991)}]{AH1991}
Hall, L.~A., \& Anderson, G.~P. 1991, Journal of Geophysical Research:
  Atmospheres, 96, 12927, \dodoi{https://doi.org/10.1029/91JD01111}

\bibitem[{Harris {et~al.}(2020)Harris, Millman, van~der Walt, Gommers,
  Virtanen, Cournapeau, Wieser, Taylor, Berg, Smith, Kern, Picus, Hoyer, van
  Kerkwijk, Brett, Haldane, del R{'{\i}}o, Wiebe, Peterson,
  G{'{e}}rard-Marchant, Sheppard, Reddy, Weckesser, Abbasi, Gohlke, \&
  Oliphant}]{numpy}
Harris, C.~R., Millman, K.~J., van~der Walt, S.~J., {et~al.} 2020, Nature, 585,
  357, \dodoi{10.1038/s41586-020-2649-2}

\bibitem[{Hartwell {et~al.}(2017)Hartwell, Knowlton, Hanson, Ennis, \&
  Maurer}]{Hartwell2017}
Hartwell, G., Knowlton, S., Hanson, J., Ennis, D., \& Maurer, D. 2017, Fusion
  Science and Technology, 72, 76, \dodoi{10.1080/15361055.2017.1291046}

\bibitem[{Haser {et~al.}(2020)Haser, Oset, \& Bodewits}]{Haser_Oset}
Haser, L., Oset, S., \& Bodewits, D. 2020, The Planetary Science Journal, 1,
  83, \dodoi{10.3847/psj/abc17b}

\bibitem[{Huebner \& Mukherjee(2015)}]{Huebner2015}
Huebner, W., \& Mukherjee, J. 2015, Planetary and Space Science, 106, 11,
  \dodoi{https://doi.org/10.1016/j.pss.2014.11.022}

\bibitem[{{Hutsem\'ekers, D.} {et~al.}(2021){Hutsem\'ekers, D.}, {Manfroid,
  J.}, {Jehin, E.}, {Opitom, C.}, \& {Moulane, Y.}}]{Opitom2021_hyak}
{Hutsem\'ekers, D.}, {Manfroid, J.}, {Jehin, E.}, {Opitom, C.}, \& {Moulane,
  Y.} 2021, A\&A, 652, L1, \dodoi{10.1051/0004-6361/202141554}

\bibitem[{Indrajith {et~al.}(2019)Indrajith, Rousseau, Huber, Nicolafrancesco,
  Domaracka, Grygoryeva, Nag, Sedmidubská, Fedor, \&
  Kočišek}]{Indrajith2019}
Indrajith, S., Rousseau, P., Huber, B.~A., {et~al.} 2019, The Journal of
  Physical Chemistry C, 123, 10639, \dodoi{10.1021/acs.jpcc.9b00289}

\bibitem[{Jessberger {et~al.}(1988)Jessberger, Christoforidis, \&
  Kissel}]{Jessberger1988}
Jessberger, E.~K., Christoforidis, A., \& Kissel, J. 1988, Nature, 332, 691,
  \dodoi{10.1038/332691a0}

\bibitem[{Johnson {et~al.}(2019)Johnson, Ennis, Loch, Hartwell, Maurer, Allen,
  Victor, Samuell, Abrams, Unterberg, \& Smyth}]{Johnson2019}
Johnson, C.~A., Ennis, D.~A., Loch, S.~D., {et~al.} 2019, Plasma Physics and
  Controlled Fusion, 61, 095006, \dodoi{10.1088/1361-6587/ab2b25}

\bibitem[{Jones {et~al.}(2017)Jones, Knight, Battams, Boice, Brown, Giordano,
  Raymond, Snodgrass, Steckloff, Weissman, Fitzsimmons, Lisse, Opitom, Birkett,
  Bzowski, Decock, Mann, Ramanjooloo, \& McCauley}]{Jones2017}
Jones, G.~H., Knight, M.~M., Battams, K., {et~al.} 2017, Space Science Reviews,
  214, 20, \dodoi{10.1007/s11214-017-0446-5}

\bibitem[{Kerkeni {et~al.}(2019)Kerkeni, Aquino, Berman, \&
  Hase}]{Boutheina2019}
Kerkeni, B., Aquino, A. J.~A., Berman, M.~R., \& Hase, W.~L. 2019, Molecular
  Physics, 117, 1392, \dodoi{10.1080/00268976.2018.1552800}

\bibitem[{Kim {et~al.}(2003)Kim, A'Hearn, Wellnitz, Meier, \& Lee}]{Kim2003}
Kim, S.~J., A'Hearn, M., Wellnitz, D., Meier, R., \& Lee, Y. 2003, Icarus, 166,
  157 , \dodoi{https://doi.org/10.1016/j.icarus.2003.07.003}

\bibitem[{Klotz {et~al.}(1996)Klotz, Marty, Boissel, Caro, Serra, Mascetti,
  Parseval, Derouault, Daudey, \& Chaudret}]{Klotz1996}
Klotz, A., Marty, P., Boissel, P., {et~al.} 1996, Planetary and Space Science,
  44, 957, \dodoi{10.1016/0032-0633(96)00026-8}

\bibitem[{Kotzian {et~al.}(1989)Kotzian, Roesch, Schroeder, \&
  Zerner}]{Kotzian1989}
Kotzian, M., Roesch, N., Schroeder, H., \& Zerner, M.~C. 1989, Journal of the
  American Chemical Society, 111, 7687, \dodoi{10.1021/ja00202a004}

\bibitem[{Kova{\v{c}}evi{\'{c}} {et~al.}(2010)Kova{\v{c}}evi{\'{c}},
  Popovi{\'{c}}, \& Dimitrijevi{\'{c}}}]{Kova2010}
Kova{\v{c}}evi{\'{c}}, J., Popovi{\'{c}}, L.~{\v{C}}., \& Dimitrijevi{\'{c}},
  M.~S. 2010, The Astrophysical Journal Supplement Series, 189, 15,
  \dodoi{10.1088/0067-0049/189/1/15}

\bibitem[{Kramida {et~al.}(2020)Kramida, {Yu.~Ralchenko}, Reader, \& {and NIST
  ASD Team}}]{NIST_ASD}
Kramida, A., {Yu.~Ralchenko}, Reader, J., \& {and NIST ASD Team}. 2020, {NIST
  Atomic Spectra Database (ver. 5.8), [Online]. Available:
  {\tt{https://physics.nist.gov/asd}} [2016, January 31]. National Institute of
  Standards and Technology, Gaithersburg, MD.}

\bibitem[{{Kurucz} {et~al.}(1984){Kurucz}, {Furenlid}, {Brault}, \&
  {Testerman}}]{Kurucz1984}
{Kurucz}, R.~L., {Furenlid}, I., {Brault}, J., \& {Testerman}, L. 1984, {Solar
  flux atlas from 296 to 1300 nm}

\bibitem[{Leadbeater(1999)}]{Leadbeater1999}
Leadbeater, N. 1999, Coordination Chemistry Reviews, 188, 35,
  \dodoi{10.1016/s0010-8545(98)00217-3}

\bibitem[{Manfroid {et~al.}(2021)Manfroid, Hutsem{\'e}kers, \&
  Jehin}]{Manfroid2021}
Manfroid, J., Hutsem{\'e}kers, D., \& Jehin, E. 2021, Nature, 593, 372,
  \dodoi{10.1038/s41586-021-03435-0}

\bibitem[{McKay {et~al.}(2019)McKay, DiSanti, Kelley, Knight, Womack,
  Wierzchos, Pinto, Bonev, Villanueva, Russo, Cochran, Biver, Bauer, Ronald
  J.~Vervack, Gibb, Roth, \& Kawakita}]{McKay2019}
McKay, A.~J., DiSanti, M.~A., Kelley, M. S.~P., {et~al.} 2019, The Astronomical
  Journal, 158, 128, \dodoi{10.3847/1538-3881/ab32e4}

\bibitem[{{Meier} {et~al.}(1998){Meier}, {Wellnitz}, {Kim}, \&
  {A'Hearn}}]{Meier1998}
{Meier}, R., {Wellnitz}, D., {Kim}, S.~J., \& {A'Hearn}, M.~F. 1998, Icarus,
  136, 268, \dodoi{10.1006/icar.1998.6022}

\bibitem[{Mommert {et~al.}(2019)Mommert, p.~Kelley, de~Val-Borro, Li, Guzman,
  Sipőcz, Ďurech, Granvik, Grundy, Moskovitz, Penttilä, \&
  Samarasinha}]{sbpy}
Mommert, M., p.~Kelley, M.~S., de~Val-Borro, M., {et~al.} 2019, Journal of Open
  Source Software, 4, 1426, \dodoi{10.21105/joss.01426}

\bibitem[{Morton(2000)}]{Morton2000}
Morton, D.~C. 2000, The Astrophysical Journal Supplement Series, 130, 403,
  \dodoi{10.1086/317349}

\bibitem[{Myllys {et~al.}(2019)Myllys, Henri, Galand, Heritier, Gilet,
  Goldstein, Eriksson, Johansson, \& Deca}]{Myllys2019}
Myllys, M., Henri, P., Galand, M., {et~al.} 2019, A\&A, 630, A42,
  \dodoi{10.1051/0004-6361/201834964}

\bibitem[{Nierenberg {et~al.}(2019)Nierenberg, Gilman, Treu, Brammer, Birrer,
  Moustakas, Agnello, Anguita, Fassnacht, Motta, Peter, \&
  Sluse}]{Nierenberg2019}
Nierenberg, A.~M., Gilman, D., Treu, T., {et~al.} 2019, Monthly Notices of the
  Royal Astronomical Society, 492, 5314, \dodoi{10.1093/mnras/stz3588}

\bibitem[{{Pollack} {et~al.}(1994){Pollack}, {Hollenbach}, {Beckwith},
  {Simonelli}, {Roush}, \& {Fong}}]{Pollack1994}
{Pollack}, J.~B., {Hollenbach}, D., {Beckwith}, S., {et~al.} 1994, \apj, 421,
  615, \dodoi{10.1086/173677}

\bibitem[{{Preston}(1967)}]{Preston1967}
{Preston}, G.~W. 1967, \apj, 147, 718, \dodoi{10.1086/149049}

\bibitem[{{Protopapa} {et~al.}(2014){Protopapa}, {Sunshine}, {Feaga}, {Kelley},
  {A'Hearn}, {Farnham}, {Groussin}, {Besse}, {Merlin}, \& {Li}}]{Protopapa2014}
{Protopapa}, S., {Sunshine}, J.~M., {Feaga}, L.~M., {et~al.} 2014, \icarus,
  238, 191, \dodoi{10.1016/j.icarus.2014.04.008}

\bibitem[{Ribar {et~al.}(2015)Ribar, Danko, Orsz{\'a}gh, Ferreira~da Silva,
  Utke, \& Matej{\v{c}}{\'i}k}]{Ribar2015}
Ribar, A., Danko, M., Orsz{\'a}gh, J., {et~al.} 2015, The European Physical
  Journal D, 69, 117, \dodoi{10.1140/epjd/e2015-50755-x}

\bibitem[{Roos-Serote {et~al.}(1995)Roos-Serote, Barucci, Crovisier, Drossart,
  Fulchignoni, Lecacheux, \& Roques}]{Roos1995}
Roos-Serote, M., Barucci, A., Crovisier, J., {et~al.} 1995, Geophysical
  Research Letters, 22, 1621, \dodoi{https://doi.org/10.1029/95GL00809}

\bibitem[{Rubin {et~al.}(2019)Rubin, Altwegg, Balsiger, Berthelier, Combi,
  De Keyser, Drozdovskaya, Fiethe, Fuselier, Gasc, Gombosi, Hänni, Hansen,
  Mall, Rème, Schroeder, Schuhmann, Sémon, Waite, Wampfler, \&
  Wurz}]{Rubin2019}
Rubin, M., Altwegg, K., Balsiger, H., {et~al.} 2019, Monthly Notices of the
  Royal Astronomical Society, 489, 594, \dodoi{10.1093/mnras/stz2086}

\bibitem[{{Schleicher} \& {A'Hearn}(1988)}]{Schleicher1988}
{Schleicher}, D.~G., \& {A'Hearn}, M.~F. 1988, \apj, 331, 1058,
  \dodoi{10.1086/166622}

\bibitem[{Schleicher \& Osip(2002)}]{Schleicher2002}
Schleicher, D.~G., \& Osip, D.~J. 2002, Icarus, 159, 210,
  \dodoi{https://doi.org/10.1006/icar.2002.6875}

\bibitem[{{Slaughter}(1969)}]{Slaughter1969}
{Slaughter}, C.~D. 1969, \aj, 74, 929, \dodoi{10.1086/110884}

\bibitem[{Str{\o}m {et~al.}(2020)Str{\o}m, Bodewits, Knight, Kiefer, Jones,
  Kral, Matr{\`{a}}, Bodman, Capria, Cleeves, Fitzsimmons, Haghighipour,
  Harrison, Iglesias, Kama, Linnartz, Majumdar, de~Mooij, Milam, Opitom,
  Rebollido, Rogers, Snodgrass, Sousa-Silva, Xu, Lin, \& Zieba}]{Strom2020}
Str{\o}m, P.~A., Bodewits, D., Knight, M.~M., {et~al.} 2020, Publications of
  the Astronomical Society of the Pacific, 132, 101001,
  \dodoi{10.1088/1538-3873/aba6a0}

\bibitem[{Sunshine \& Feaga(2021)}]{Sunshine2021}
Sunshine, J.~M., \& Feaga, L.~M. 2021, The Planetary Science Journal, 2, 92,
  \dodoi{10.3847/psj/abf11f}

\bibitem[{Villanueva {et~al.}(2018)Villanueva, Smith, Protopapa, Faggi, \&
  Mandell}]{PSG}
Villanueva, G., Smith, M., Protopapa, S., Faggi, S., \& Mandell, A. 2018,
  Journal of Quantitative Spectroscopy and Radiative Transfer, 217, 86,
  \dodoi{https://doi.org/10.1016/j.jqsrt.2018.05.023}

\bibitem[{Virtanen {et~al.}(2020)Virtanen, Gommers, Oliphant, Haberland, Reddy,
  Cournapeau, Burovski, Peterson, Weckesser, Bright, {van der Walt}, Brett,
  Wilson, Millman, Mayorov, Nelson, Jones, Kern, Larson, Carey, Polat, Feng,
  Moore, {VanderPlas}, Laxalde, Perktold, Cimrman, Henriksen, Quintero, Harris,
  Archibald, Ribeiro, Pedregosa, {van Mulbregt}, \& {SciPy 1.0
  Contributors}}]{scipy}
Virtanen, P., Gommers, R., Oliphant, T.~E., {et~al.} 2020, Nature Methods, 17,
  261, \dodoi{10.1038/s41592-019-0686-2}

\bibitem[{Wyckoff {et~al.}(1999)Wyckoff, Heyd, \& Fox}]{Wyckoff1999}
Wyckoff, S., Heyd, R.~S., \& Fox, R. 1999, Astrophysical Journal, 512, L73,
  \dodoi{10.1086/311869}

\bibitem[{Öberg {et~al.}(2017)Öberg, Guzm{\'{a}}n, Merchantz, Qi, Andrews,
  Cleeves, Huang, Loomis, Wilner, Brinch, \& Hogerheijde}]{Berg2017}
Öberg, K.~I., Guzm{\'{a}}n, V.~V., Merchantz, C.~J., {et~al.} 2017, The
  Astrophysical Journal, 839, 43, \dodoi{10.3847/1538-4357/aa689a}

\end{thebibliography}
\appendix
\section{Fluorescence Efficiencies of Observed Metal Lines}\label{app:fluor}

\begin{figure*}[b]
\centering
\includegraphics[width=0.95\textwidth]{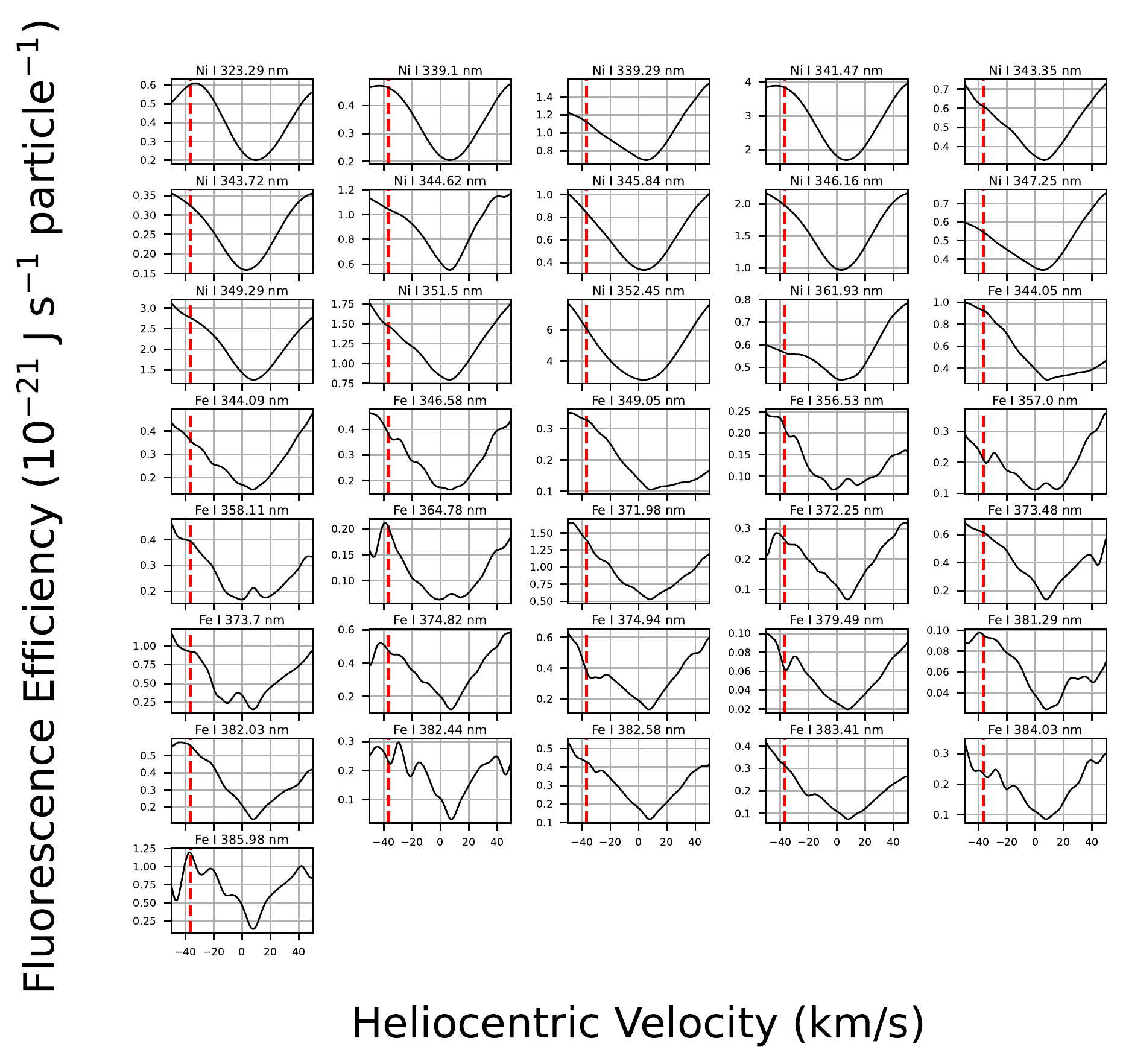}\caption{Calculated fluorescence efficiencies for observed metal lines in Hyakutake as a function of heliocentric velocity. The velocity of Hyakutake (-36.7~km/s) is indicated by a vertical red line in each plot. {Each spectra is shown with independent linear intensity scale.}}\label{fig:fluor_eff_all}
\end{figure*}

\end{document}